\documentclass{ieeeaccess}
\usepackage{cite}
\usepackage{amsmath,amssymb,amsfonts}
\usepackage{algorithmic}
\usepackage{graphicx}
\usepackage{textcomp}
\usepackage{mathtools}
\usepackage{subfigure}
\usepackage{float} %Added to fix figure positions. @astro
\usepackage{hyperref} % For clickable references and citations
\usepackage{color,soul} %For Highlighting and review
\usepackage{listings} %For code highlighting

\newcommand\floor[1]{\lfloor#1\rfloor}

\def\BibTeX{{\rm B\kern-.05em{\sc i\kern-.025em b}\kern-.08em
    T\kern-.1667em\lower.7ex\hbox{E}\kern-.125emX}}

\begin{document}
\history{This work has been submitted to the IEEE Access for possible publication. Copyright may be transferred without notice, after which this version may no longer be accessible.}
\doi{10.1109/ACCESS.2021.DOI}

\title{Random Access Procedure over Non-Terrestrial Networks:\\ From Theory to Practice}
\author{\uppercase{Oltjon Kodheli}, \uppercase{Abdelrahman Astro, Jorge Querol, Mohammad GHOLAMIAN, Sumit Kumar, Nicola Maturo, and Symeon Chatzinotas}}
\address{Snt - Interdisciplinary Centre for Security,
	Reliability and Trust, University of
	Luxembourg, 29 Avenue J.F.
	Kennedy,Luxembourg City L-1855,
	Luxembourg}

\tfootnote{This work has been partially supported by the Luxembourg National Research Fund (FNR) under the Industrial Fellowship project SATIOT (12526592), and the CORE project MegaLEO (C20/IS/14767486).}

\markboth
{Oltjon Kodheli \headeretal: Random Access Procedure over Non-Terrestrial Networks: From Theory to Practice}
{Oltjon Kodheli \headeretal: Random Access Procedure over Non-Terrestrial Networks: From Theory to Practice}

\corresp{Corresponding author: Oltjon Kodheli (e-mail: oltjon.kodheli@uni.lu).}

\begin{abstract}
Non-terrestrial Networks (NTNs) have become an appealing concept over the last few years and they are foreseen as a cornerstone for the next generations of mobile communication systems. Despite opening up new market opportunities and use cases for the future, the novel impairments caused by the signal propagation over the NTN channel, compromises several procedures of the current cellular standards. One of the first and most important procedures impacted is the random access (RA) procedure, which is mainly utilized for achieving uplink synchronization among users in several standards, such as the fourth and fifth generation of mobile communication (4 \& 5G) and narrowband internet of things (NB-IoT). In this work, we analyse the challenges imposed by the considerably increased delay in the communication link on the RA procedure and propose new solutions to overcome those challenges. A trade-off analysis of various solutions is provided taking into account also the already existing ones in the literature. In order to broaden the scope of applicability, we keep the analysis general targeting 4G, 5G and NB-IoT systems since the RA procedure is quasi-identical among these technologies. Last but not least, we go one step further and validate our techniques in an experimental setup, consisting of a user and a base station implemented in open air interface (OAI), and an NTN channel implemented in hardware that emulates the signal propagation delay. The laboratory test-bed built in this work, not only enables us to validate various solutions, but also plays a crucial role in identifying novel challenges not previously treated in the literature. Finally, an important key performance indicator (KPI) of the RA procedure over NTN is shown, which is the time that a single user requires to establish a connection with the base station.
\end{abstract}

\titlepgskip=-15pt
\maketitle
%%%%%%%%%%%%%%%%%%%%%%%%%%%%%%%%%%%%%%%%%%%%%%%%%%%%%%%%%%%%%%%%%%%%%%%%%%%%%%%%%%%%%%

\section{Introduction}\label{sec1}

Orthogonal frequency division multiple access (OFDMA) is an efficient multiple-access technique utilized in modern wireless communication systems including NB-IoT, 4G long term evolution (LTE) and 5G new radio (NR) \cite{ofdma}. The most important precondition for these technologies is establishing a time-frequency synchronization between the users and the base station. The synchronization in downlink is achieved by specific synchronization signals that are broadcasted by the base station to all the users in specific time-intervals, whereas the uplink synchronization is obtained by an explicit procedure known as random access (RA) procedure. A failure in the synchronization phase, either downlink or uplink, would deprive the users form accessing the network and being able to transmit or receive data. 

Non-terrestrial networks, as defined by the 3rd Generation Partnership Project (3GPP), are a popular concept nowadays for current and near-future generations of mobile communications, due to their unique capabilities in providing connectivity in areas where a terrestrial infrastructure is unfeasible or cost-inefficient. In fact, a work item (WI) has already started in order to enable NTN-based 5G systems \cite{tr38821}, and a study item (SI) is about to conclude regarding NB-IoT via NTN \cite{tr36763}. Nevertheless, despite from bringing several important advantages and opening up new market opportunities, the presence of the NTN channel, being quite different from a terrestrial one, imposes new challenges to be carefully considered \cite{5gntnsurvey}. One of these challenges is the increased delay in the communication link \cite{fraunhofer, survey}, and the first phase that will be directly impacted is the uplink synchronization enabled by the RA procedure. Notably, depending on the altitude and the type of the NTN platform the delay will vary, reaching round trip time (RTT) values of up to 480 ms for a GEO satellite with transparent payload. 

There exist several works in the literature that have analyzed the impact of the increased propagation delay over the NTN channel on the RA procedure. One of the well known and widely accepted solution to cancel out this increased delay is the application of a Timing Advance (TA) from the users when transiting their uplink signal to the base station \cite{ericsson-5gntn, ericsson-iotntn, kodheli-5gntn}. According to these works, a Global Navigation Satellite System (GNSS) receiver can be used in order to estimate the location of the serving satellite and the user itself, so as to derive the TA required. Other works put forward the idea of a fixed TA employed only by taking into account the delay experienced at the center of the satellite antenna footprint (reported to the users by the base station), while leaving the other part to be handled in the same manner as in a terrestrial scenario \cite{harri-5gntn, guidotti-5gntn}. The differential delay part (with reference to the center of the antenna footprint) would limit the supported beam size by the protocol, as derived in \cite{guidotti-beamsize}. To extend the supported beam size, new preamble formats for the RA procedure should be specifically designed targeting NTN systems, as proposed in \cite{prdesign-5gntn}. Another widely accepted solution by all previous works is  the increase of the relevant timers in the RA procedure that would force its failure. Obviously, the proposed timers greatly depend on the altitude of the NTN terminal and the protocol considered. In fact, although the RA procedure is quite similar in LTE, NR and NB-IoT, the corresponding timers related to this process may change from one standard to another. It is worth highlighting here that these challenges and solutions have been also treated in 3GPP, mainly summarized in \cite{tr38821, tr36763}. 

While all the above-mentioned proposals to counteract the impact of the increased delay on the RA procedure hold in theory, some of them fail in practice according to the findings of this paper. This is because none of the current works have done an in-depth analysis of the protocol, either LTE, NR or NB-IoT in the context of NTN operation. In addition, validation of the current proposals is missing and other crucial challenges that would lead to a failure in the RA procedure over NTN have been ignored. To cover this gap in the literature, we do an in-depth analysis of the RA procedure over NTN and the corresponding challenges imposed by the increased signal propagation delay. We propose novel solutions to counteract the impact of the increased delay on the RA procedure and do a trade-off analysis of different approaches, taking into account also the already existing ones in the literature. Last but not least, we go one step further and validate our proposed solutions on a realistic experimental setup utilizing the Open Air Interface (OAI) implementations of the 3GPP users and base station and an NTN channel emulator implemented in hardware (HW). The experimental setup has played a crucial role in this work because it allowed us to identify through a trail-and-error approach the new challenges not previously treated in the literature, and to test various solutions from a practical point of view. Overall, the contributions of this work can be summarized as follows:

\begin{itemize}
    \item We identify new challenges, not treated so far, regarding the RA procedure over NTN imposed by the considerable increased delay in the communication link. 
    \item We propose novel solutions for the challenges identified in this paper, as well as for the ones already treated in other works, and provide a trade-off analysis of various approaches.
    \item In order to test our solutions, we design a testbed based on OAI implementation for the 3GPP users and base station, and HW implementation for the NTN channel emulating the experienced delay in the communication link between the users and the base station.
    \item We measure and show an important KPI of the RA procedure over NTN, such as the single user access time. This will act as a lower bound for future works in this area where more than one user can access the network and collisions may occur during the RA procedure.
\end{itemize}
It is worth emphasizing here that for the purpose of broadening the scope of applicability, we keep the analysis of the RA procedure general targeting LTE, NR and NB-IoT standard. As previously mentioned, this important procedure for achieving uplink synchronization is quasi-identical among these protocols, as we will see later on. 

The paper is organized as follows. In Section \ref{sec2} we provide a brief overview of the RA procedure in cellular networks. In Section \ref{sec3} we analyse the challenges of the RA procedure when adopted over NTN and in Section \ref{sec4} we propose techniques to counteract the challenges and provide a trade-off analysis among various approaches. Section \ref{sec5} describes the laboratory setup of our experiment and explains the implementation of the proposed algorithms. Our solutions are validated and the user access time under different NTN delays is shown. Finally, we conclude the paper in Section \ref{sec6} by summarizing the work and highlighting the future directions.

\Figure[t!](topskip=0pt, botskip=0pt, midskip=0pt)[width=120mm,keepaspectratio]{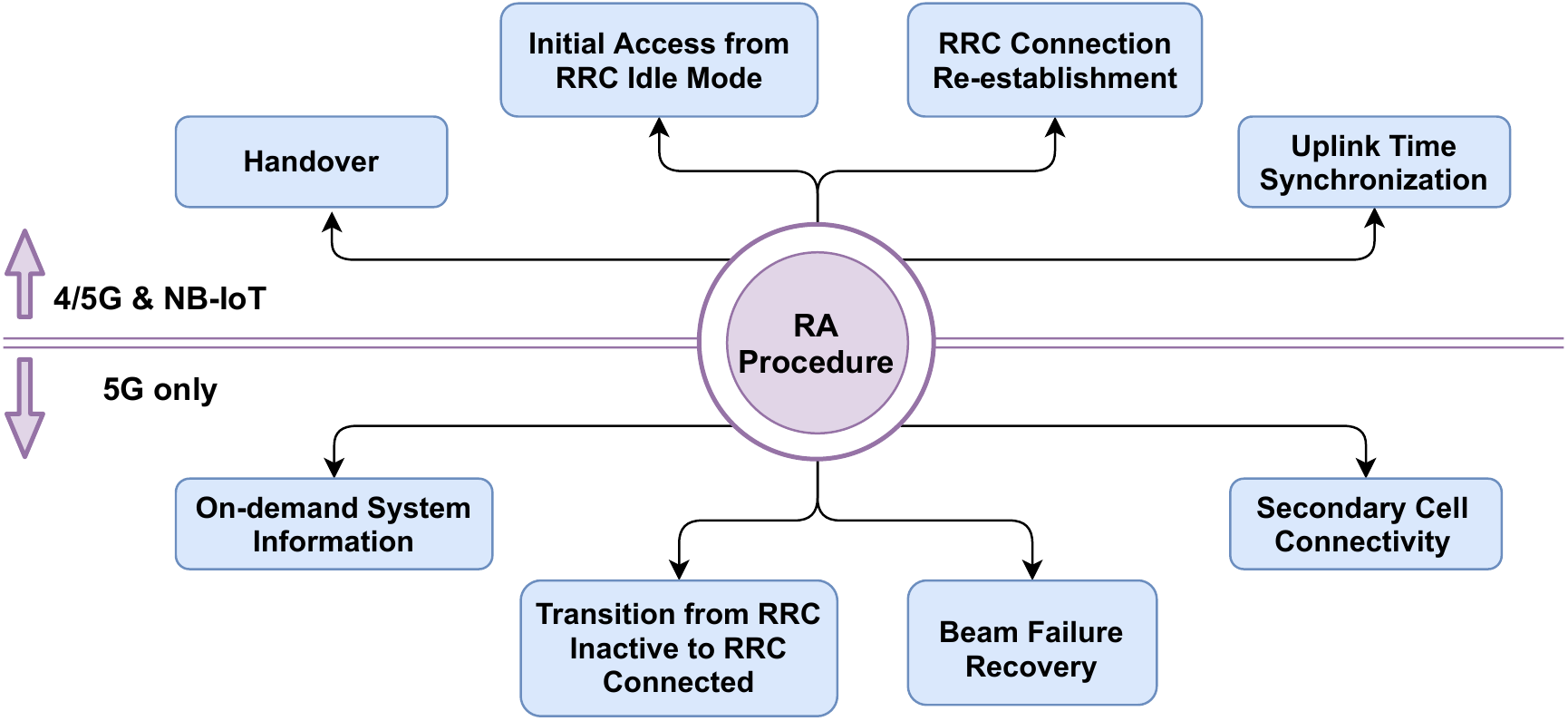}
{Situations when RA procedure is triggered.\label{fig1}}
%Figure a UE Side, we may add another state correspondent to the max HARQ.

%%%%%%%%%%%%%%%%%%%%%%%%%%%%%%%%%%%%%%%%%%%%%%%%%%%%%%%%%%%%%%%%%%%%%%%%%%%%%%%%%%%%%%%%

\section{A brief overview of the RA procedure in cellular networks}\label{sec2}

The downlink synchronization of a cellular network is achieved by specific signals broadcasted to all the user equipments (UEs) by the base station (BS), termed as primary synchronization signal (PSS) and secondary synchronization signals (SSS). If applied in the uplink, this broadcasting/ always-on synchronization mechanism would not be efficient due to the limited power at the user side and the high amount of interference it would be generated. Therefore, an alternative more efficient procedure is utilized, known as the RA procedure. There are two types of RA, contention-based and contention-free. While contention-based RA consists of a 4-step message exchange between the UEs and the BS, the contention-free mechanism uses only a 2-step message exchange. The RA procedure is triggered in different situations as illustrated in Figure \ref{fig1}. Please note that compared to LTE and NB-IoT, there is a stronger necessity of the RA procedure in NR, and this is also an indicator of why this mechanism is highly important, and worth analysing its feasibility over NTN. A more detailed description of each steps involved in the RA procedure is given below. 

%\begin{table}[!b] \caption{List of Common Acronyms}  %\label{list_of_acr}
%\centering
% \begin{tabular}{|l|l| }
% \hline
% \textbf{Acronyms} & \textbf{Definitions} \\ 
% \hline
% 3GPP & The 3rd Generation Partnership Project  \\
% NTN & The Non-Terrestrial Networks  \\
% RA & Random Access  \\
% NB-IoT & Narrowband Internet of Things  \\
% KPI & Key Performance Indicator \\
% OAI & Open Air Interface  \\
% LTE & Long Term Evolution \\
% NR & New Radio \\
% TA & Time Advance \\
% TD & Time Delay \\
% RTT & Round Trip Time \\
% SF & Subframe Number \\
% UE & User Equipment \\
% BS & Base Station \\
% PRACH & Physical Random Access Channel \\
% GEO & Geostationary Earth Orbit \\
% GNSS & Global Navigation Satellite System \\
% PSS & Primary Synchronization Signal \\
% SSS & Secondary Synchronization Signal \\
% DMRS & Demodulation Reference Signals \\
% LEO & Low Earth Orbit \\
% PHY & Physical Layer \\
% MAC & Medium Access Control \\

% \hline
% \end{tabular}
%\end{table}
\subsection{Message 1}

After achieving downlink synchronization, the UEs are able to decode important control channels containing useful information for the subsequent steps, such as the configuration index of physical random access channel (PRACH). The "channel" terminology in LTE, NB-IoT or NR is used to identify the type of data that is carried over specific time-frequency resources. For example, the Message 1 is sent only over the PRACH channel and the time-frequency resources of the PRACH, which is also known as the RA opportunity (RAO), are derived by the configuration index. When there is a RAO, the UE will send a random access preamble to the serving BS, by using the PRACH. This allows the BS to estimate the RTT for each UE, based on the time of arrival (ToA) of the received preamble. The BS will utilize the ToA estimate for determining a timing advance (TA) to be applied by each UE. Doing so, the uplink transmission from various UEs will be time-synchronized. What is worth emphasizing here is that at this step the UEs will compete for the same PRACH channel (same RAO), hence packet collision may occur. To reduce the probability of collision, there exist a list of possible preamble sequences defined in the standard and the UEs randomly select one. Notably, in case two UEs randomly select the same preamble sequence to initiate the RA procedure, a collision will occur, leading to a failure of the RA procedure. In such a case, the UEs will try again to send the preamble after a back-off time and a power-ramping.
\begin{figure}[!b]
	\centering
	\includegraphics[width=75mm,keepaspectratio]{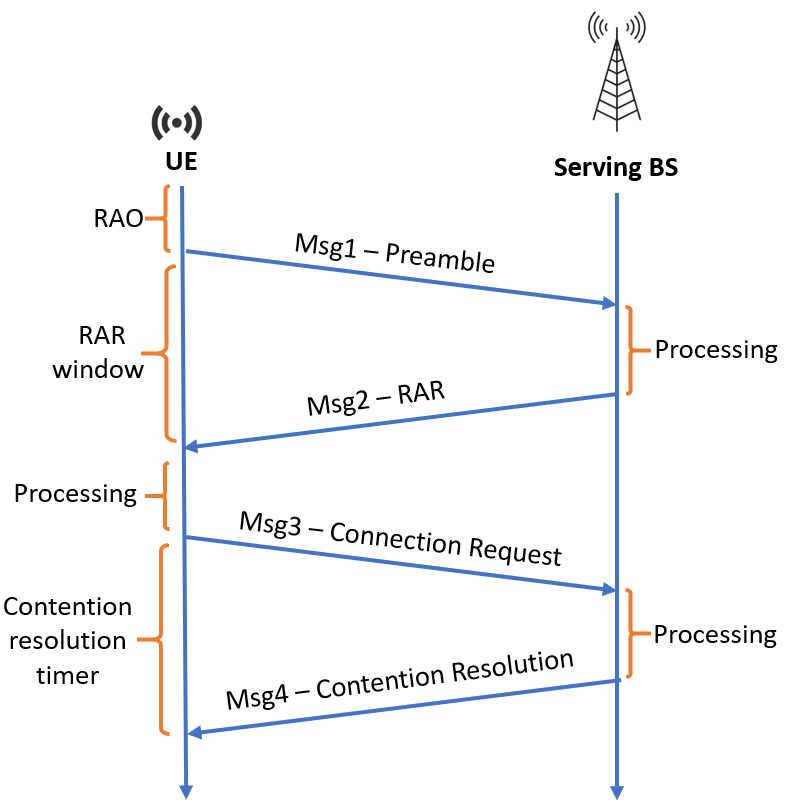}
	\caption{A graphical representation of the RA procedure}
	\label{fig2}
\end{figure}
 \begin{figure*}[!t]
\begin{centering}
\subfigure a){\includegraphics[width=80mm,keepaspectratio]{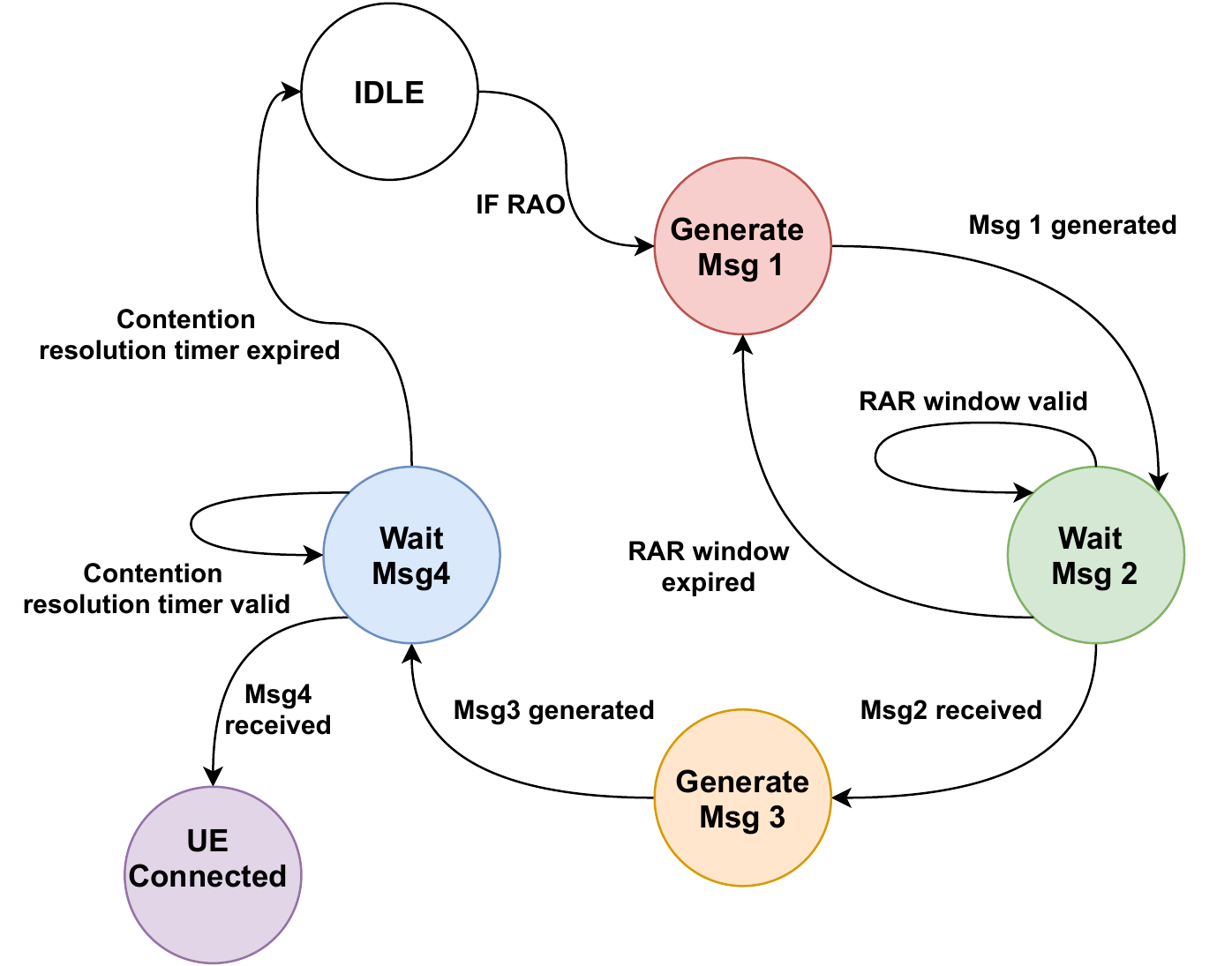}}
\subfigure b){\includegraphics[width=80mm,keepaspectratio]{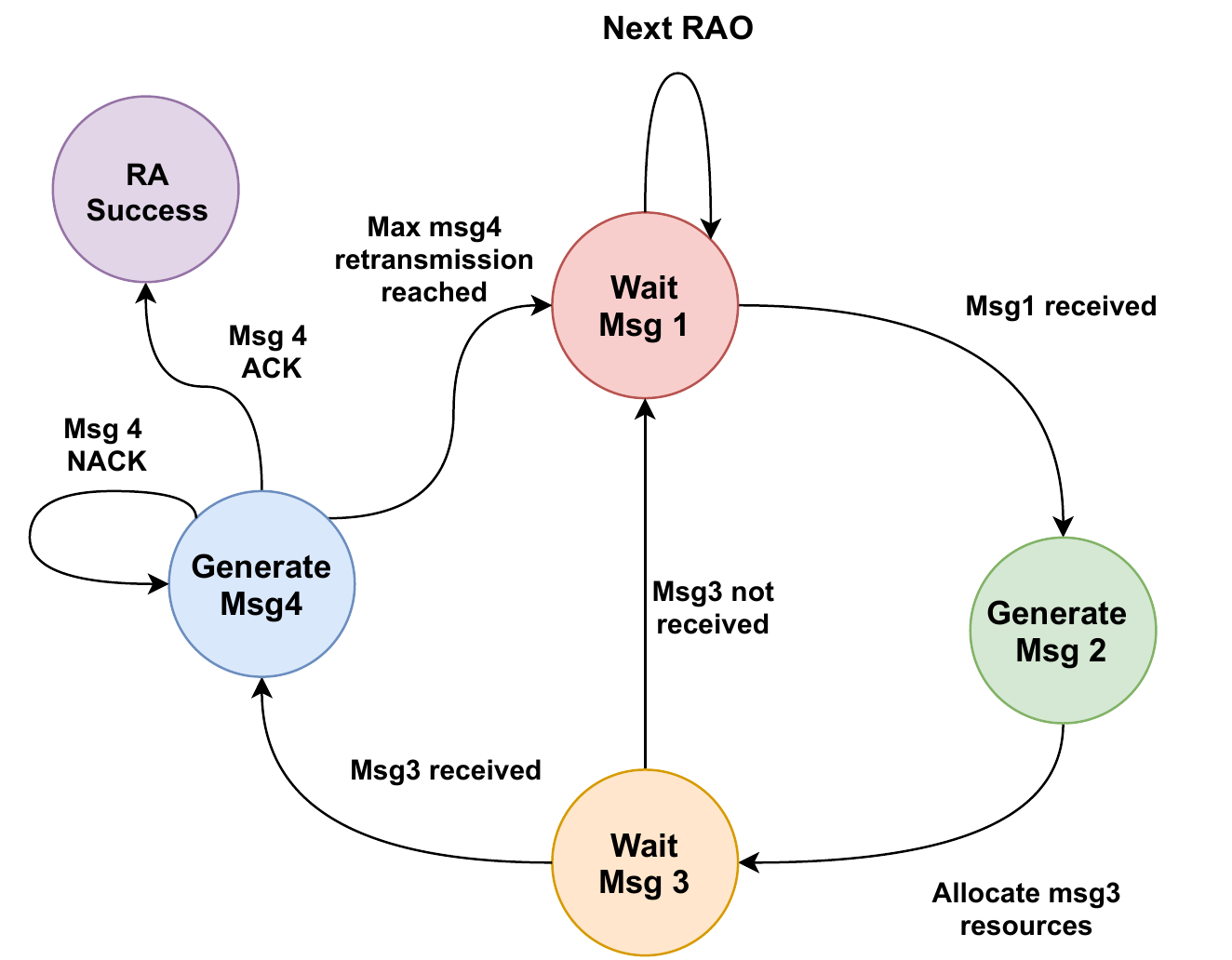}}
\caption{State machine representation during the RA procedure at: a) UE side and b) BS side.}
\label{fig3}
\par\end{centering}
\end{figure*}
\subsection{Message 2}

The BS will continuously check for preamble reception at a RAO and in case it detects one, it will respond with a random access response (RAR) known as Message 2. The RAR contains the TA parameter that we mentioned earlier, as well as the scheduling information pointing to the radio resources that the UEs has to utilize for subsequent uplink data transmission and the modulation and coding scheme (MCS). All the future message exchanges between the UEs and the BS, either in downlink or uplink, are now orchestrated by the BS. In contrast to Message 1 transmission, collision avoidance is achieved by assigning distinct time-frequency resources to different UEs. Last but not least, the UEs wait for a limited amount of time for Message 2 reception, as defined in the standard. This is also known as the RAR window size. If the RAR is not received inside a certain window size, the RA procedure is considered as unsuccessful. Again, the RA procedure has to be repeated according to the rules defined in the standard.

\subsection{Message 3}

Until this stage, the UE does not have a unique identity (ID) in the network. The RA preamble sequence number cannot act as an ID because the same one can be selected by multiple UEs. Therefore, the main goal of Message 3 transmission is to initiate a connection request where the UE is introduced in the network with a unique ID (known as C-RNTI). This phase is also known as the contention resolution phase. Please note that in case of a contention-free RA Message 3 and Message 4 transmission are skipped because in such situations the user is already uniquely identified (e.g. when synchronization is lost). Together with a randomly selected C-RNTI, the UE may also report its data volume status and power headroom to facilitate the scheduling and power allocation algorithms for subsequent transmissions.

\subsection{Message 4}

In this final step, the BS will send back to the UE the confirmation regarding the selected temporary C-RNTI, which will act as a permanent ID for the user for all the future message exchanges. Similar to Message 2 reception, also in this case the UE will wait for Message 4 until the contention resolution timer is valid. If this timer expires, the UE will re-attempt the RA procedure again at another RAO. Last but not least, hybrid automatic repeat request (HARQ) protocol is adopted for Message 3 and 4 transmission. It mainly consists of an extra message indicating the reception or not (ACK or NACK) of a certain packet. In case of NACK, the same packet has to be re-transmitted. Figure \ref{fig2} illustrates the four steps involved in the RA procedure, whereas Figure \ref{fig3} shows the state machine representation at the UE and BS side, in an example scenario where the UE is attempting the initial access, transitioning from radio resource control (RRC) Idle mode to RRC Connected mode. This will facilitate and assist the explanation of the challenges in the following section.
 
%%%%%%%%%%%%%%%%%%%%%%%%%%%%%%%%%%%%%%%%%%%%%%%%%%%%%%%%%%%%%%%%%%%%%%%%%%%%%%%%%%%%%%%

\begin{figure*}[!t]
\begin{centering}
\subfigure a){\includegraphics[width=65mm,keepaspectratio]{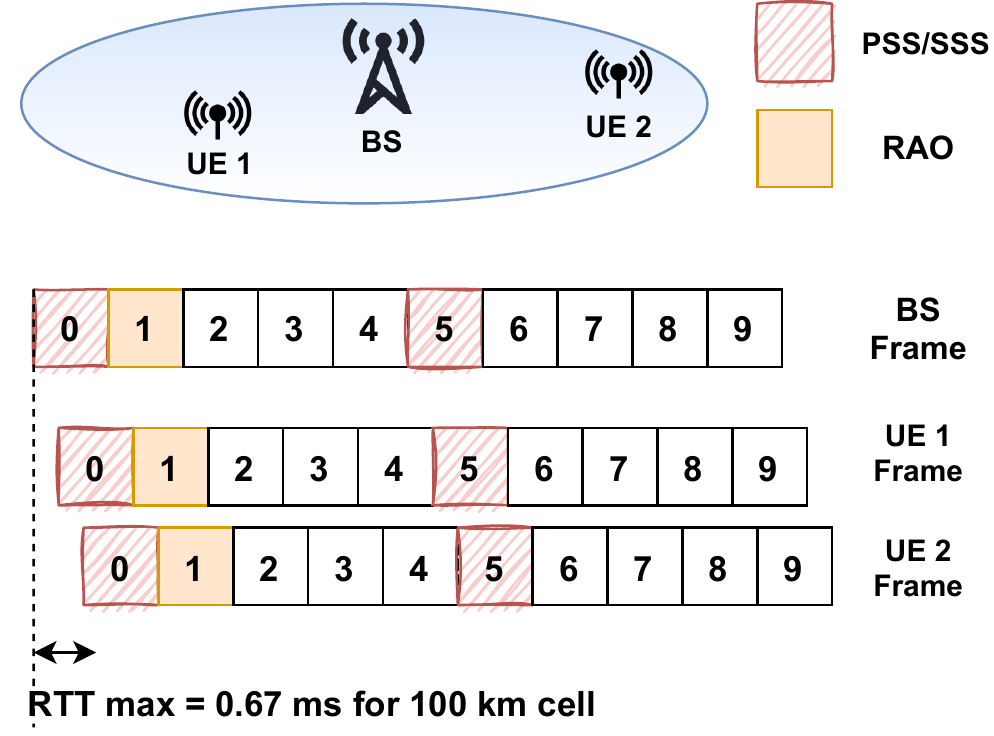}}
\subfigure b){\includegraphics[width=73mm,keepaspectratio]{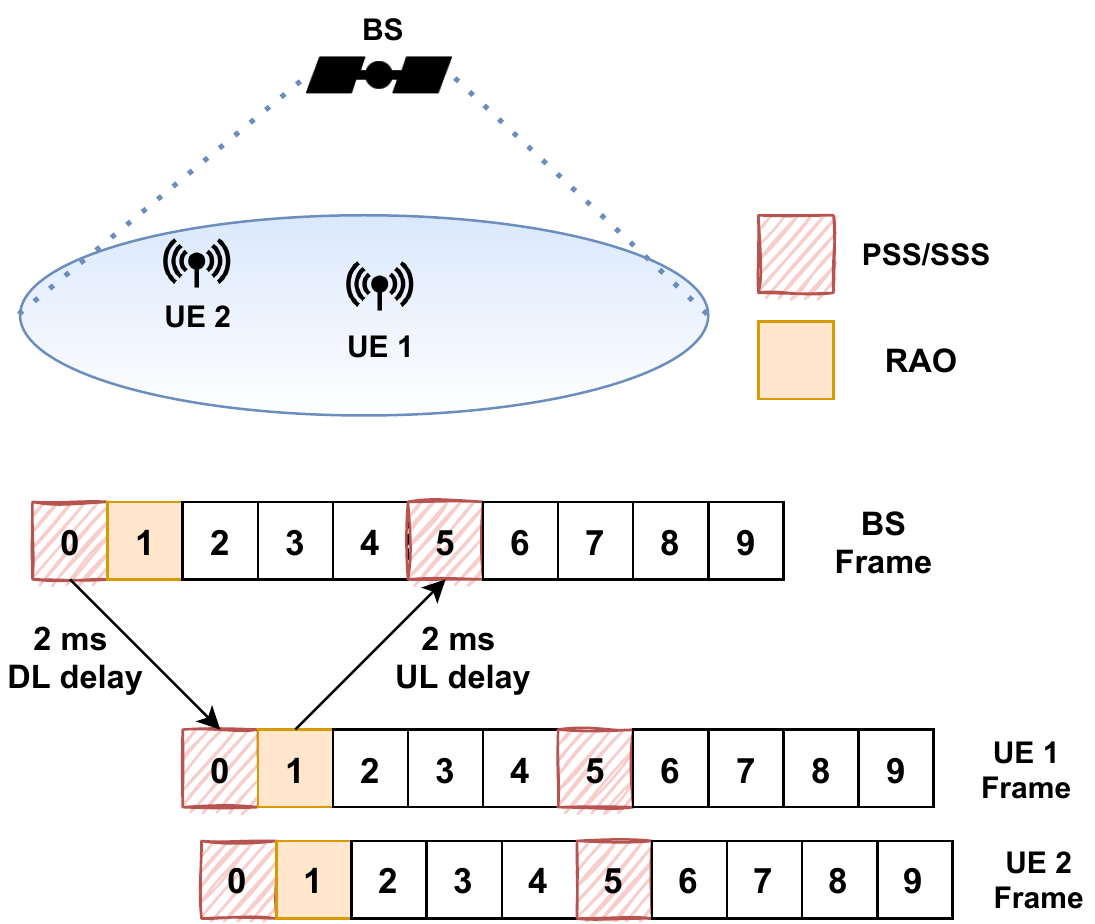}}
\caption{LTE scenario example: a) terrestrial and b) NTN - satellite at 600 km altitude.}
\label{fig4}
\par\end{centering}
\end{figure*}
\section{Challenges of RA Procedure over NTN}\label{sec3}

The aim of this section is to analyse the obstacles that would make the RA procedure fail when larger delays (compared to terrestrial scenario) are introduced in the communication link. Before going on with the investigation, we would firstly like to emphasize another crucial aspect that is common among LTE, NR and NB-IoT. The radio frames of all these three important technologies are 10 ms long, consisting of 10 subframes (SF), each of 1 ms duration. This shared feature enables us to keep the analysis general since the following challenges would hold for all these three technologies. To improve the material flow, the technical divergences among them are highlighted when needed.
\begin{table}[b]%
	\caption{PRACH configuration index mapped into RAO time \& periodicity \cite{ts36211, ts38211}. \label{tab1}}
	\centering
	\begin{tabular*}{240pt}{@{\extracolsep\fill}|c||cc|cc|@{\extracolsep\fill}}
		
		\hline
		\multicolumn{1}{|c||}{\textbf{Configuration}} &
		\multicolumn{2}{c|}{\textbf{LTE}} &
		\multicolumn{2}{c|}{\textbf{NR}} \\
	    \cline{2-5}
		{\textbf{Index}}& SF Number & Frame & SF Number & Frame \\
		\hline
		0 & 1 & Even & 1 & 1, 17, 33,..\\
		
		1 & 4 & Even & 4 & 1, 17, 33,.. \\
		
		2 & 7 & Even & 7 & 1, 17, 33,.. \\
		
		3 & 1 & Any & 9 & 1, 17, 33,.. \\

	4 & 4 & Any & 1 & 1, 9, 17,... \\
	
		- & - & - & - & - \\
		
	- & - & - & - & - \\

		\hline
	\end{tabular*}
\end{table}

\subsection{RAO mismatch between the UE \& BS} \label{sec3a}

The PSS/SSS signals that are responsible for downlink synchronization, are sent over fixed SF, known by the UEs. This allows the UEs under the coverage area of the BS to be synchronized at a frame and SF level, as well as assign the SF numbering with reference to the PSS/SSS placement inside the frame. On the other hand, the system frame number is included in the master information block (MIB), which is directly decoded by the UEs after PSS/SSS synchronization process. As previously stated, the RAO time and periodicity is determined by the PRACH configuration index, which is encapsulated and broadcasted to all the UEs through the system information block (SIB). Table \ref{tab1} gives only a set of possible RAO time/periodicity for LTE and NR. Obviously, a much larger number of configurations exist, especially for NR, which can be specifically found in the standard (Table 5.7.1-2 for LTE \cite{ts36211} and Table 6.3.3.2-2 for NR \cite{ts38211}). In NB-IoT there is no PRACH configuration index reported, thus the PRACH time and periodicity are communicated separately in SIB (Section 10.1.6 \cite{ts36211}). However, the principle would be the same. Assuming an LTE BS, operating over a PRACH configuration index 3, the UEs will initiate the RA procedure at SF number 1 at any frame. Also the BS will check at SF 1 for preamble detection and measure the ToA of the preambles, so as to estimate the TA and correct the time-misalignment (see Figure \ref{fig4}.a).

Let us now assume the same BS operating over an NTN platform (e.g. LEO satellite at 600 km altitude). The minimum RTT in such a scenario would be 4 ms. Due to this drastic change of the delay, the PSS/SSS signals will arrive at the UE at least 2 ms after they are produced. Nevertheless, despite the increased delay, the SF-level synchronization and numbering at the downlink is not impacted, only shifted in time (2 SF in our example as shown in Figure \ref{fig4}.b). This time-shift in downlink, together with the time that the signal required to propagate in the uplink, would cause a mismatch between the RAO at the UE and BS. Referring to the state machines in Figure \ref{fig3}, the BS will be stuck forever on state "Wait Msg1" if no modifications are done in the current standard, because it will process only SF 1 for preamble reception according to this PRACH configuration index, while in fact the preambles will arrive at SF 5.

\subsection{Message 2 withdrawal by the UE}\label{sec3b}

Assuming that we are able to solve the above-mentioned challenge, the next obstacle that would occur in the RA procedure is the Message 2 withdrawal by the UEs.  When the UE initiate the RA procedure, it assigns to the preamble a specific random access ID, known as RA-RNTI, which is explicitly derived by the time-frequency resources utilized for preamble transmission. The following equations are utilized for RA-RNTI derivation, as specified in the standard (NB-IoT/LTE \cite{ts36321}, NR \cite{ts38321}).
\begin{equation}
\text{\textbf{LTE:}} \hspace{0.5cm}   RA-RNTI = 1 + t_{id} + 10 \cdot f_{id}
\end{equation}
\begin{equation}
\text{\textbf{NB-IoT:}} \hspace{0.2cm}   RA-RNTI=1 + \floor{\frac{t_{id}}{4}} + 256 \cdot c_{id} 
\end{equation}
\begin{equation}
\text{\textbf{NR:}} \hspace{0.05cm}   RA-RNTI=1 + s_{id} + 14 \cdot t_{id} + 14 \cdot 80 \cdot f_{id} + 14 \cdot 80 \cdot 8 \cdot c_{id}
\end{equation}
Please note that $t_{id}$ is the index of the SF related to the RAO when the preamble is sent, $f_{id}$ is the index that relates to the frequency used for preamble transmission, $c_{id}$ is the carrier index and $s_{id}$ is the symbol index. The exact same equations are utilized also by the BS to calculate the RA-RNTI that will be encoded in the RAR message, which will then be sent to the UE. However, similar to the previous problem, because there is a mismatch in the SF numbering of the RAO between the UE and the BS, the RA-RNTI associated with Message 1 transmission would be different from the one received together with Message 2. The UE will withdraw the RAR message received because it will assume that it belongs to another user. In fact, also the parameters related to the frequency domain may be impacted due to the frequency shift caused by the high-speed movement of the NTN platform. However, this analysis is out of the scope of this work, and we study only the impact of the timing difference. In such a case, the UE will be stuck on the loop "Wait Msg2 - Generate Msg1" according to the state machines shown in Figure \ref{fig3}, because the RAR window size will always expire.

\subsection{Expiry of the RAR window size and contention resolution timer}

Even if we are able to solve the Message 2 withdrawal problem, for certain values of NTN altitudes the RAR window size will expire. When this happens, the UEs assume that their preamble is lost and will repeat sending Message 1 after a back-off time and a power ramping procedure, as specified by the standard. The same holds for the contention resolution (CR) phase. When the UE Transmit Message 3 to the BS, it will monitor the downlink channels for Message 4 reception until the expiration of the CR timer. The set of values for the $RAR_{window}$ and $CR_{timer}$ for NB-IoT/LTE \cite{ts36331} and NR \cite{ts38331} are given below:
\begin{equation} \label{eq4}
\begin{split}
\text{\textbf{LTE:}} \hspace{0.05cm} RAR_{window} &= \{1, 2, 4, 6, 8, 10\} \,\, SF \\
 CR_{timer} & = \{8, 16, 24, 32, 40, 48, 56, 64\} \,\, SF
\end{split}
\end{equation}
\begin{equation} \label{eq5}
\begin{split}
\text{\textbf{NB-IoT:}} \hspace{0.05cm} RAR_{window} &= \{2, 3, 4, 5, 6, 7, 8, 10\} \,\, PP \\
 CR_{timer} & = \{1, 2, 3, 4, 8, 16, 32, 64\} \,\, PP
\end{split}
\end{equation}
\begin{equation} \label{eq6}
\begin{split}
\text{\textbf{NR:}} \hspace{0.05cm} RAR_{window} &= \{1, 2, 4, 8, 10, 20, 40, 80\} \,\, SL \\
 CR_{timer} & = \{8, 16, 24, 32, 40, 48, 56, 64\} \,\, SF
\end{split}
\end{equation}
Please note that SF stands for subframe duration, SL for slot duration and PP for physical downlink control channel (PDCCH) periodicity. As it can be seen, the maximum RAR window and CR timer supported in LTE are 10 ms and 64 ms, respectively. For NB-IoT, the calculations are done based on PDCCH periodicity, which varies from 1 ms to 10.24 seconds. Notably, these range of timers are enough to cover even NTN delays, which in the worst case scenario would be as high as 480 ms (GEO satellite with transparent payload). Hence, this challenge does not apply to NB-IoT. Last but not least, the calculations for the NR are a bit more complex because of its flexibility in the PHY layer. While it is true that the frame structure is the same as NB-IoT/LTE, the number of slots contained in 1 SF and the slot duration have a direct dependency on the subcarrier spacing (SCS). The NR standard \cite{ts38211} defines several SCS for the OFDM signal, thus in Table \ref{tab2} we show the maximum RAR window supported for every numerology. Finally, regarding the CR timer for NR, it is identical to LTE. 

\begin{table}[b]%
	\caption{Slot duration and maximum RAR window sizes for various NR numerology. \label{tab2}}
	\centering
	\begin{tabular*}{210pt}{@{\extracolsep\fill}|c||c|c|c|@{\extracolsep\fill}}
 \hline
		{\textbf{$\mu$}}& $ SCS = 2^\mu \cdot 15$ & SL duration & Max $RAR_{window}$\\
		\hline
		0 & 15 kHz & 1 ms & 80 ms\\
		
		1 & 30 kHz & 0.5 ms & 40 ms\\
		
		2 & 60 kHz & 0.25 ms & 20 ms \\
		
		3 & 120 kHz & 0.125 ms & 10 ms \\

	4 & 240 kHz & 0.0625 ms & 5 ms\\
		\hline
	\end{tabular*}
\end{table}

\subsection{Message 3 \& HARQ scheduling mismatch between the UE \& BS} \label{sec3d}

\begin{figure}[!t]
	\centering
	\includegraphics[width=80mm,keepaspectratio]{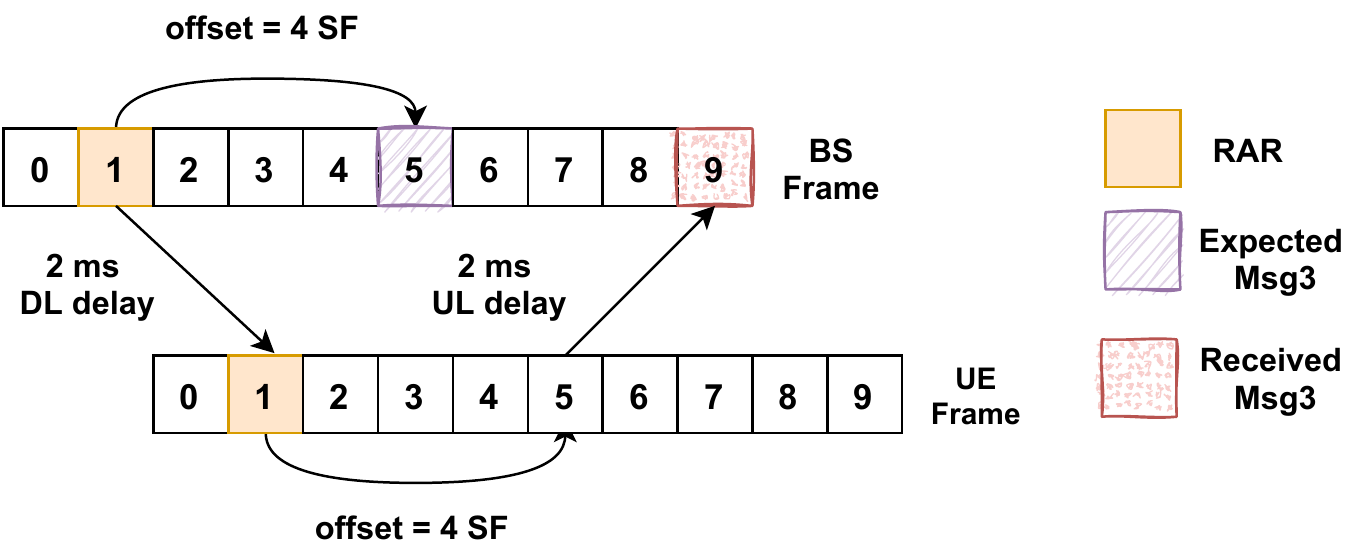}
	\caption{Scheduling problem illustration (e.g. satellite at 600 km altitude)}
	\label{fig5}
\end{figure}

The CR phase of the RA procedure is orchestrated by the BS, and the time-frequency resources that the UE has to utilize for Message 3 are reported in the RAR. Apparently, this is done to avoid collision among different UEs. In the time-domain, the resource allocation for Message 3 is done through informing the UEs about the amount of time they have to wait (time-offset) after RAR reception, before transmitting their uplink signals. 
The minimum value for this time-offset is 4 SF (ms) in NB-IoT/LTE \cite{ts36213} because this corresponds also to the amount of processing time at the user side before initiating an uplink transmission. Regarding NR, the minimum value of such time-offset is dependent on the numerology \cite{ts38213}. Based on the offset that the UE has to apply, also the BS has to prepare for Message 3 reception. However, due the the increased delay in the communication link (which overcomes the SF length and the BS is not aware of in the current protocol), Message 3 will arrive at the BS not according to the resource allocation (see Figure \ref{fig5}). This will cause a failure in Message 3 reception and the BS will never enter in the "Generate Msg4" state (Figure \ref{fig3}). The same reasoning can be done also for the HARQ scheduling. The BS will monitor certain SF for receiving the ACK/NACK during the CR phase, which in fact will be further delayed.

\subsection{Message 3 demodulation \& decoding problems}\label{sec3e}

Solving the scheduling problem investigated earlier, does not still guarantee a correct reception of Message 3 at the BS side. This is because of the following reasons.

\subsubsection{Demodulation reference signals (DMRS) mismatch}

For the purpose of coherent demodulation and channel estimation, both at UE and BS side, reference symbols (pilots) are inserted in the downlink and uplink channels. Message 3 is sent over the physical uplink shared channel (PUSCH), therefore hereafter we will provide more details regarding the DMRS signals for the PUSCH channel to better explain this challenge. For NB-IoT/LTE standard, DMRS occupies the center symbol of every slot (contained by 7 symbols overall). Although understanding the positioning of DMRS sequence is quite straightforward, its generation is a bit more complex and involves many parameters. Referring to the NB-IoT/LTE standard \cite{ts36211}, the generation of DMRS sequence directly depends on three main parameters as shown below: 
\begin{equation}
    r_{DMRS} (n) \sim e^{j\alpha n} \cdot r_{u,v}(n)
\end{equation}
where $\alpha$ is the cyclic shift which defines the rotation of the sequence, $r_{u,v}$ is the base sequence determined by two other parameters (UE-specific) $u$ and $v$ that define the group hopping and sequence hopping, respectively. The cyclic shift $\alpha$ is given by:
\begin{equation}
    \alpha = 2 \pi \cdot \frac{\{ n_{DMRS}^{(1)} + n_{DMRS}^{(2)}  + n_{PRS}(SF) \}mod12}{12}
\end{equation}
where $ n_{DMRS}^{(1)}$ and $ n_{DMRS}^{(2)}$ take integer values from 0 to 10 defined in the standard \cite{ts36211}. What is worth noticing here is the last parameter that has a direct dependency on the SF number. This results in applying different cyclic shifts to the base sequence, leading to distinct DMRS symbols inserted at every SF. To correctly demodulate Message 3, the BS will have to utilize the same symbols (pilots), allowing to extract relevant parameters for the demodulation. However, due to the the SF-mismatch caused by the increased delay that overcomes the SF length (as shown in Figure \ref{fig5}), the estimated parameters for the demodulation process will be erroneous, greatly distorting Message 3 reception. Please note that we provided further details for NB-IoT/LTE to assist the explanation of this challenge, but the same will occur also in NR because of the SF-dependency of the DMRS sequence \cite{ts38211}.

\subsubsection{Gold sequence mismatch}

The Gold sequences are used in various wireless technologies as a scrambling code because of their high auto-correlation and low cross-correlation properties. Through the scrambling code that is applied to the bits after encoding and before proceeding with the modulation, the BS can separate signals coming simultaneously from various UEs,  as well as the UEs can distinguish the signals coming from many different BS. In NB-IoT/LTE \cite{ts36211} and NR \cite{ts38211}, the Gold sequence $c$ is generated by combining two m-sequences $x_1$ and $x_2$ through an XOR gate, as follows:
\begin{equation}
    c(n) = \{x_1(n +N_c) + x_2 (n + N_c)\}mod2
\end{equation}
where $N_c = 1600$ is fixed. The final obtained Gold sequence will depend on the initialization value:
\begin{equation}
    c_{init} = x_{1init} + x_{2init}(SF)
\end{equation}
As it can be noted, the initialization of one of the m-sequences has a SF-dependency, resulting in distinct scrambling codes for different SF. This will cause Message 3 decoding issues because the de-scrambling sequence applied at the BS side will be different from the scrambling one at the UE (similar to the demodulation problem previously explained).

Please note that the demodulation and decoding problems will occur only in the uplink transmission (UE -> BS). In the downlink (BS -> UE), the SF numbering matches, being assigned with reference to PSS/SSS location in the frame, as previously emphasized.

%%%%%%%%%%%%%%%%%%%%%%%%%%%%%%%%%%%%%%%%%%%%%%%%%%%%%%%%%%%%%%%%%%%%%%%%%%%%%%%%%%%%%%%%

\section{Proposed Solutions to overcome the challenges and trade-off analysis}\label{sec4}

The aim of this section is to describe the proposed solutions which counteract the challenges treated in Section \ref{sec3}. Furthermore, a trade-off analysis will be provided for the various approaches.
\begin{table}[b]%
	\caption{PRACH configuration index mapped into RAO time \& periodicity at UE side for NR over NTN scenario. \label{tab3}}
	\centering
	\begin{tabular*}{195pt}{@{\extracolsep\fill}|c||cc|@{\extracolsep\fill}}
		
		\hline
		\multicolumn{1}{|c||}{\textbf{Configuration}} &
		\multicolumn{2}{c|}{\textbf{NR over NTN}} \\
	
	    \cline{2-3}
		{\textbf{Index}}& SF Number & Frame \\
		\hline
		0  & 1 - SFA & 1-FA, 17-FA, 33-FA,..\\
		
		1  & 4 - SFA & 1-FA, 17-FA, 33-FA,.. \\
		
		2  & 7 - SFA & 1-FA, 17-FA, 33-FA,.. \\
		
		3  & 9 - SFA & 1-FA, 17-FA, 33-FA,.. \\

	4  & 1 - SFA & 1-FA, 17-FA, 33-FA,.. \\
	
		-  & - & - \\
		
	-  & - & - \\

		\hline
	\end{tabular*}
\end{table}

\subsection{Timing advance (TA) at the UE side} \label{sec4a}

To compensate for the RAO mismatch between the UE and BS, a TA can be applied, so that the preambles transmitted by the on-ground UEs arrive at the right RAO at the BS, as specified by the PRACH configuration parameters. To do so, the UEs has to estimate the RTT in the NTN channel, which can be realized through the GNSS positioning capability together with the knowledge of the NTN orbital parameters. Although this solution is already proposed in the literature \cite{ericsson-5gntn, ericsson-iotntn, kodheli-5gntn}, it lacks a detailed description implementation-wise. The existing TA concept, either in NB-IoT, LTE or NR, can be only applied in a sample-level, and due to this, there are practical limitation to be considered. To illustrate this, let us take as an example the TA application in LTE. The minimum TA that can be applied in LTE is $TA =  16 \cdot T_s \cdot TAC$, where $T_s$ is the sample duration given by:
\begin{equation}
    T_s = \frac{1}{FFT_{size} \cdot SCS} = \frac{1}{2048 \cdot 15000}
\end{equation}
and the TAC is the timing advance command reported in RAR. There are 11 bits reserved for reporting TAC, ranging from 0 to 1282, corresponding to TA values of up to 0.67 ms. Although being able to estimate higher values of the TA before transmitting Message 1, it is practically impossible to apply a TAC greater than 2048 samples which is also the FFT size in LTE. The TAC needed to counteract NTN delays has to be much larger than the SF length, thus overcoming the FFT size. Alternative ways should be proposed.

In this work, we put forward the idea of applying a SF-level TA at the UE side, in addition to the sample-level one already defined by the standard. 
This can be done through applying a time correction in the mapping of the PRACH configuration index into RAO time and periodicity at the UE side, as shown in Table \ref{tab1}, while leaving it unchanged at the BS. 
Table \ref{tab3} gives an example of how the configuration index would be mapped in RAO time and periodicity for NR. Obviously, the specific SF advance (SFA) and/or frame advance (FA) will have a direct dependency on the old value of the RAO Frame and SF, and the estimated RTT. For example, if the BS is operating over an PRACH configuration index 3 and the estimated RTT by a specific UE is 4.3 ms, SFA will be 4 and FA will be 0. The remaining part of 0.3 ms can be handled by the sample-level TA. 

In a similar way, relying on GNSS estimates at the UE side, such SF-timing correction can be also included in Equation (1) - (3) to solve the Message 2 withdrawal by the UE, and in Equation (7) - (10) to solve the demodulation and decoding issues. Doing so would make the RA-RNTI associated with the sent preamble to match the one received in RAR, and enable correct  demodulation/decoding at the BS.

\subsection{SF-level timing delay (TD) at the BS side} \label{sec4b}

\begin{figure}[!t]
	\centering
	\includegraphics[width=80mm,keepaspectratio]{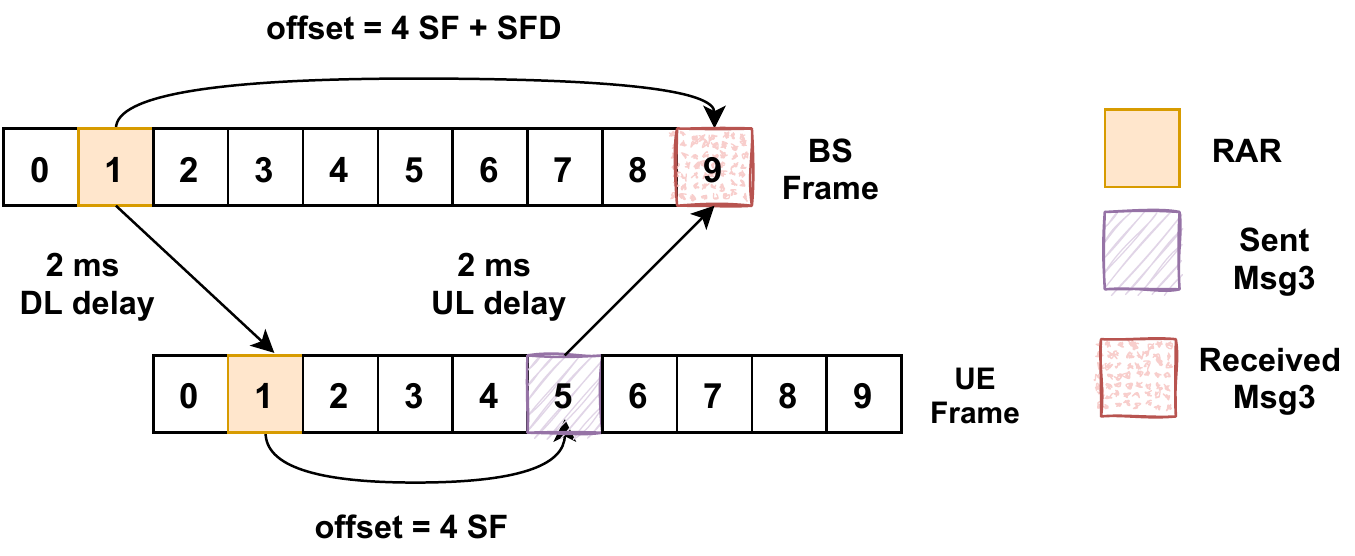}
	\caption{Solving the scheduling problem by adding the SF Delay (SFD)}
	\label{fig6}
\end{figure}

The TA concept, even at a SF-level, fails to solve the scheduling mismatch between the UE and the BS treated in Section \ref{sec3d}. As we explained, the resource allocation in time-domain for Message 3 is performed by assigning a time-offset to the UEs, which has to be applied after the reception of the RAR message (where the time-offset information is included). If we take the NB-IoT standard as an example, such time-offset ranges from 4 to 64 \cite{ts36213}. However, the RTT over an NTN channel overcomes this range of values, making the application of the TA totally unfeasible when the time-offset is set to 4 (required for processing), and partially unfeasible for the other set of values (depending on the NTN altitude). Therefore, this raises the need of a SF-level TD to be applied at the BS, as illustrated in Figure \ref{fig6}. Please note that in order to delay the reception of Message 3, the BS has to be aware of the NTN altitude and the RTT experienced at the center of its beam. Such information can be provided/included in the deployment phase of the network since it will be fixed over time. 

In a similar way, the SF-level TD can be executed at the BS side for the RAO mismatch. Clearly, this will require a new mapping of the PRACH configuration index into the RAO time and periodicity at the BS, resembling Table \ref{tab3}, but with the difference of applying a SF delay (SFD) and frame delay (FD) instead of SFA and FA. Furthermore, the same concept can be utilized for SF-timing corrections in Equation (1) - (3) and (7) - (10). This solution will not require any modification at the UE side. 

\begin{figure}[!b]
	\centering
	\includegraphics[width=78mm,keepaspectratio]{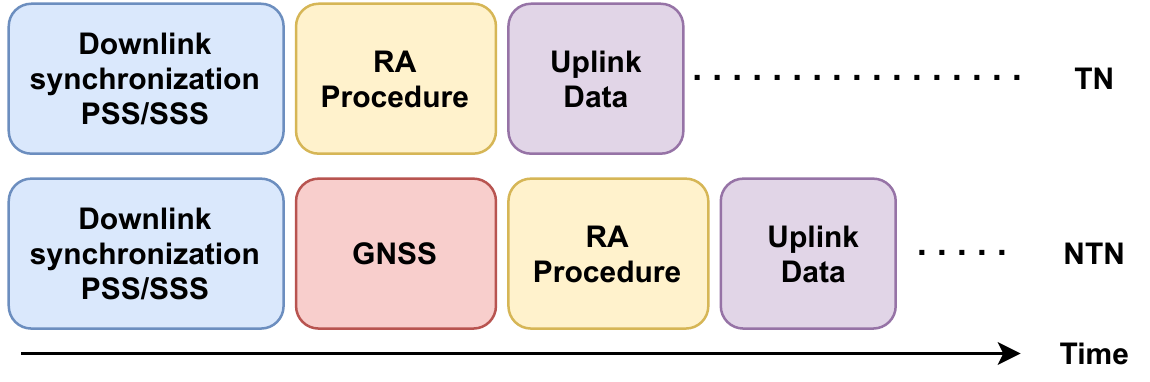}
	\caption{Chronology of UE procedures for TN and NTN}
	\label{fig7}
\end{figure}

\subsection{Adaptation of timers}
\label{sec4c}
To avoid the expiry of the RAR window size and contention resolution timer, the adaptation can be done either at the UE or the BS. If done at the UE side, again the GNSS estimates will play a role. The estimation of the RTT has to be added on top of the values reported by the BS, as shown in Equation (4) - (6). If done at the BS side, the new values should be reported in the SIB, where all the relevant parameters for the RA procedure are included. In the current NB-IoT/LTE and NR standard, only 3-bits are reserved for reporting the RAR window size, and other 3-bits for the contention resolution timer. This is because there are only up to 8 possible values reported. Including more values in the current standards with the purpose of covering also the NTN delays would require extra bits in the SIB field.

\subsection{Trade-off analysis} \label{sec4d}

As it has been pointed out, some of the above-mentioned solutions can be implemented at the UE or the BS. This raises the need for a trade-off analysis to compare among these two different approaches.

\subsubsection{Power consumption}

The TA approach covered in Section \ref{sec4a} relies on the utilization of the GNSS signals to estimate the UE position, before starting the RA procedure (see Figure \ref{fig7}). To have a rough estimation of the RTT in the NTN channel, the UE has to be provided also with the two-line element set (TLE) of the NTN platform in order to predict its location over time. This added procedure compared to a terrestrial network (TN) will directly impact the power consumption of the UE, which is a very sensitive matter, especially for the NB-IoT standard. In fact, the increased power consumption due to GNSS has been deeply analyzed in \cite{tr36763}, by considering various commercial UEs and scenarios. The simulation results provided in \cite{tr36763} show a reduction in battery life in the range of 10 - 40 $\%$ for UEs placed in a medium coverage level with maximum coupling loss (MCL) of 154 dB. Clearly, the diversity of the battery life-time reduction comes from the various applications the UEs are being used for (different packet sizes for the uplink data reports), and the various GNSS modules utilized for positioning. 

In contrast to the TA, implementing the TD approach at the BS side, as described in Section \ref{sec4b}, will eliminate the need for the extra GNSS processing at the UE (without an impact in the battery life-time), leading to coverage limitations, as we will see below.

\begin{table}[b]%
	\caption{CP length for some preamble formats. \label{tab4}}
	\centering
	\begin{tabular*}{140pt}{@{\extracolsep\fill}|c||c|c|c|@{\extracolsep\fill}}
		
		\hline
		\multicolumn{1}{|c||}{\textbf{Preamble}} &
		\multicolumn{3}{c|}{\textbf{CP length in ms}} \\
	
	    \cline{2-4}
		{\textbf{Format}}& NB-IoT & LTE & NR \\
		\hline
		0  & 0.027 & 0.1 & 0.1 \\
		
		1  & 0.067 & 0.68 & 0.68 \\
		
		2  & 0.8 & 0.2 & 0.15 \\

		\hline
	\end{tabular*}
\end{table}
\begin{figure}[!t]
	\centering
	\includegraphics[width=80mm,keepaspectratio]{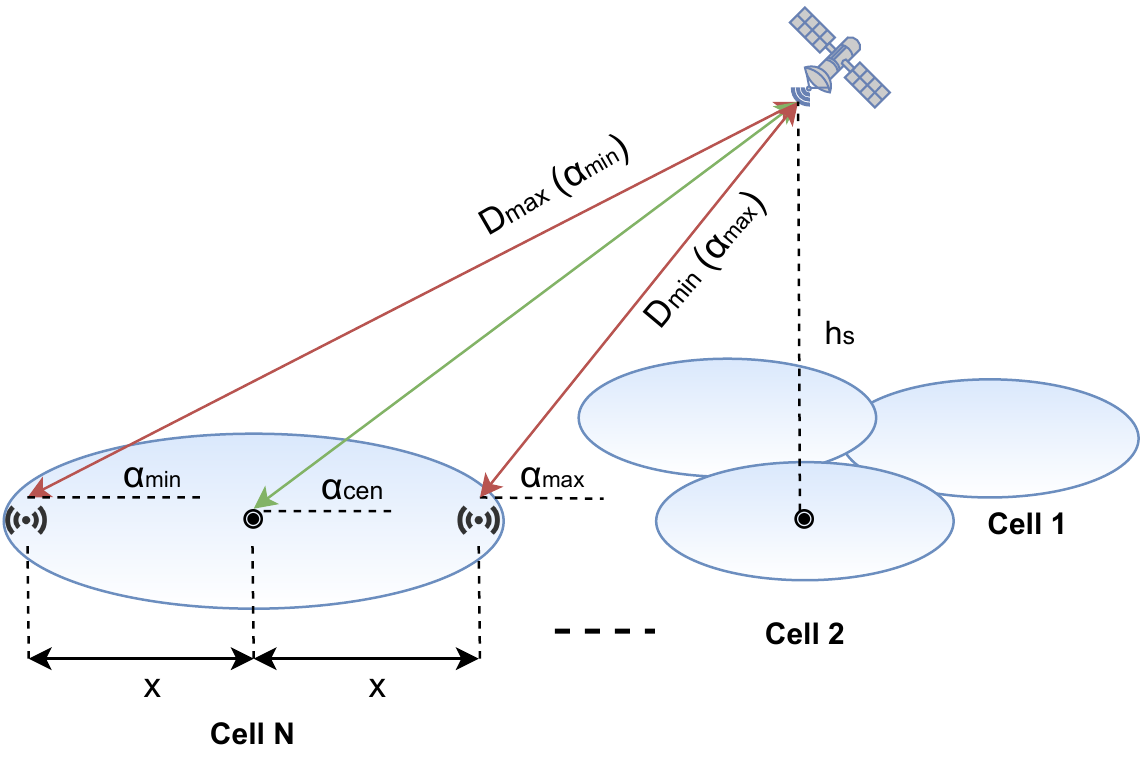}
	\caption{NTN system geometry}
	\label{fig8}
\end{figure}
\subsubsection{NTN Coverage}

In a terrestrial network, the cell radius $R_{cell}$ is dependant on the cyclic prefix (CP) length $T_{cp}$ of the preamble generated during Message 1 transmission of the RA procedure. To correctly receive the preambles from various UEs in a RAO, the differential RTT of signal propagation among the UEs should not overcome the CP length of the preamble. This can be expressed by the following equation:
\begin{equation}
   D_{max} - D_{min} =  R_{cell} \leq c \cdot T_{cp}/2
\end{equation}
where $D_{max}$ and $D_{min}$ represent the maximum and the minimum UE distance from the BS. There exist different preamble formats defined in the NB-IoT/LTE \cite{ts36211} and NR \cite{ts38211} standard, where some are captured in Table \ref{tab4}. As it can be seen, the CP length alters with the preamble format, resulting in distinct cell sizes.
\begin{figure}[!t]
	\centering
	\includegraphics[width=85mm,keepaspectratio]{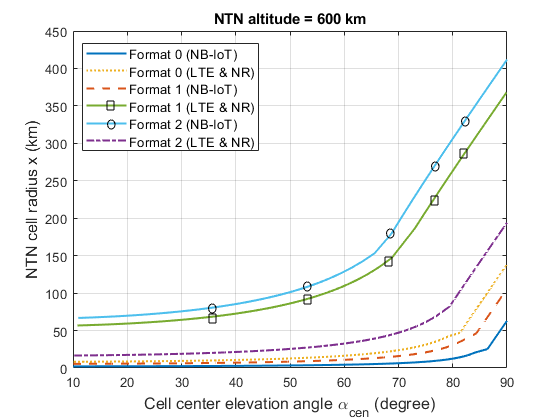}
	\caption{NTN cell radius as a function of the cell center elevation angle.}
	\label{fig9}
\end{figure}
\begin{figure*}[!t]
	\centering
	\includegraphics[width=\linewidth,keepaspectratio]{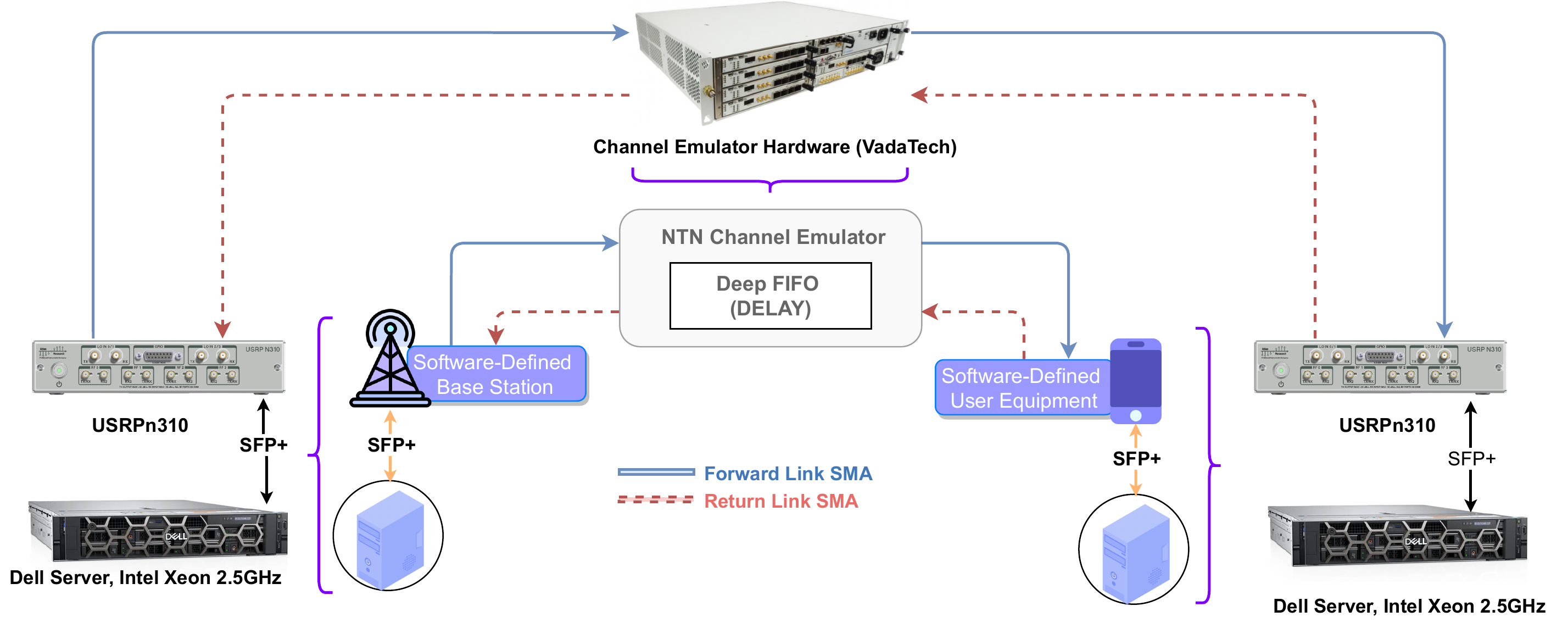}
	\caption{Infrastructure setup for the realization of Non-Terrestrial Network}
	\label{fig10}
\end{figure*}
\begin{figure*}[t]
	\centering
	\includegraphics[width=0.9\linewidth,keepaspectratio]{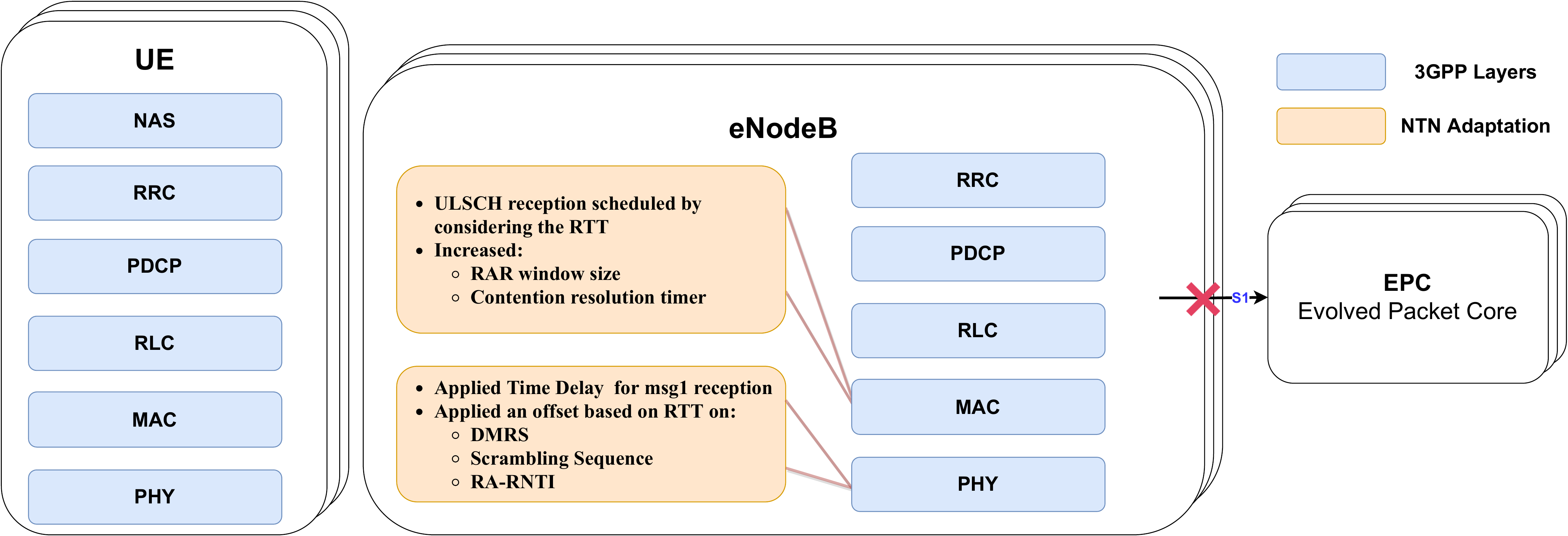}
	\caption{{OpenAirInterface LTE Stack with NTN adaptations - TD Approach}}
	\label{fig11}
\end{figure*}
While in a terrestrial network $D_{min} = 0$ and $D_{max} = R_{cell}$, the same does not hold in NTN (please refer to Figure \ref{fig4}). To derive the cell sizes in NTN we exploit the system geometry in Figure \ref{fig8} and write the following equations, in a similar way to \cite{guidotti-beamsize}. The distance $D$ from a UE to the BS in NTN can be calculated by:
\begin{equation}
D(\alpha) =\Big(\sqrt{R_E^2  sin(\alpha)^2 + h_s^2 + 2 R_E  h_s} -R_E \cdot sin(\alpha)\Big) 
\end{equation}
where $\alpha$ is the elevation angle, $h_s$ is the altitude of the NTN platform and $R_E$ is the radius of Earth. Given the minimum elevation angle $\alpha_{min}$ of a certain cell, we can easily derive $D_{max}(\alpha_{min})$ by Equation (13), and $D_{min}(\alpha_{max})$ by the following:
\begin{align}
\label{eq7}
 D_{min}(\alpha_{max}) & = \min_{\alpha} D(\alpha)\\
&\text{such that:} \notag \\
& D_{max}(\alpha_{min}) - D(\alpha) \leq c \cdot T_{cp}/2 \nonumber \\
& \alpha \in [\alpha_{min} \,\,\,\,\,\,\, 90^{\circ}] \nonumber
\end{align}
After obtaining $D_{min}$ and $D_{max}$ for a particular cell, the radius $x$ can be calculated as follows:
\begin{equation}
\label{eq8}
x = \frac{\sqrt{D_{max}^2 + D_{min}^2 + 2D_{max}D_{min} cos(\alpha_{max}+\alpha_{min})}}{2} 
\end{equation}
Through numerical simulations, we obtain the NTN cell sizes for different preamble formats in NB-IoT, LTE and NR, for a LEO satellite at 600 km altitude (see Figure \ref{fig9}). For simplicity, we show the results with respect to the cell center elevation angle $\alpha_{cen}$, derived after fixing $\alpha_{min}$ and obtaining $x$.

As it can be observed, there is a limitation in the size of the NTN cells supported, driven by the preamble format and $\alpha_{cen}$. This does not occur in the case of TA approach, because the TA is applied individually at Message 1 by the UEs, making sure that they arrive at the BS at the correct RAO. Whereas, in the TD approach, the BS can delay the RAO to accumulate the preambles from all the UEs inside a cell, but cannot have several RAO. Alternative NB-IoT/LTE or NR carriers are needed for the other cells, having different PRACH configurations and applying different TD at the BS, depending on the characteristics of the cell. Please note that the TD does not need to be continuously estimated, but rather fixed at the deployment phase of the network. For example, if the range of the RTT in a cell is 4.1 - 4.6 ms, the BS has always to apply a fixed SFD = 4 for that cell, while leaving the residual 0.1 - 0.6 ms to be handled by the protocol itself (clearly the preamble format selected should have $T_{cp}\geq 0.6$). Another interesting observation here is that certain cells cannot provide services at all. For example, let us assume an LTE over NTN scenario, and a cell where the range of RTT is between 5.6 and 5.9 ms. While the BS can apply SFD = 5, the residual part of 0.6 to 0.9 ms cannot be estimated/corrected with the current existing preamble formats. This will obviously cause further limitations in the coverage. Again, such restriction does occur in the case of the TA technique, because in addition to the SF-level TA, a sample-level TA can be employed at the UE side. 

%%%%%%%%%%%%%%%%%%%%%%%%%%%%%%%%%%%%%%%%%%%%%%%%%%%%%%%%%%%%%%%%%%%%%%%%%%%%%%%%%%%%%%%%

\begin{figure*}[t]
	\centering
	\includegraphics[width=\linewidth,keepaspectratio]{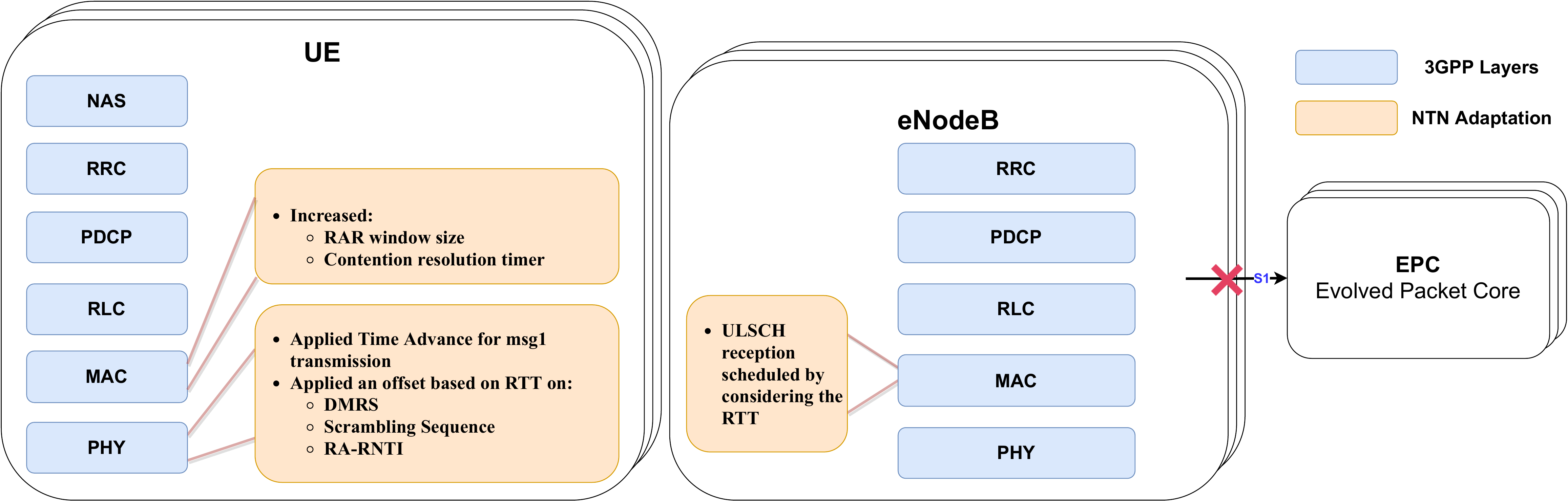}
	\caption{{OpenAirInterface LTE Stack with NTN adaptations - TA Approach}}
	\label{fig12}
\end{figure*}

\section{Experimental Validation}\label{sec5}

\subsection{Experimental Setup}

The infrastructure illustrated in Figure \ref{fig10} has been developed using OpenAirInterface (OAI) \cite{oai5g}. OAI is open-source software that is being used for the experimentation, evaluation, development of current 3GPP wireless communication standards such as LTE and 5G \cite{oai}, and several 3GPP proposals towards 5G and beyond. The experimental validation of our proposed techniques has been done by employing the LTE protocol stack since it is the most mature one currently developed by OAI. Figure \ref{fig11} and \ref{fig12} show the block diagram of OAI implementation of the 3GPP LTE stack coupled with our incorporated NTN adaptations, spanning Physical (PHY) and Medium Access Control (MAC) layers. As it can be noted, both the TA and TD approach have been tested, as described in Section \ref{sec4}. More specifically, we deployed the OAI LTE stack on general-purpose processors coupled with software-defined radio for signal generation and acquisition. This experimental setup focuses more on the radio access network (RAN) side with abstracted connection to the EPC so-called $noS1$ mode. Hence, it allows testing of OAI base-station and UEs without the support of the core network, which is a great advantage of OAI for accelerating the development time. 

The physical interconnections between the system components is also illustrated in Figure \ref{fig10}. The connection between the processing units and the software-defined radio (SDR) units is established via small form-factor pluggable (SFP+) transceivers, which support 10 Gigabit Ethernet (GigE) connection. Accordingly, it gives a better realization for high-speed communication networks between the processing units (host computers) and SDRs. Thus, the host-based OAI software can seamlessly control the SDR hardware to receive and transmit data over the $10GigE$ link.

The NTN channel emulator, a core component for verifying and validating the NTN development, has been implemented on top of a field-programmable gate array (FPGA) using custom-designed hardware from VadaTech. It provides an IP-Core implementing the RTT delay using a deep first in first out (FIFO) buffer, which utilizes an external Double Data Rate (DDR) Synchronous Dynamic Random-Access Memory (SDRAM) chipset. The implemented delay length is based on the frequency rate at which the data writes and reads to/from the Deep FIFO, and the depth of the FIFO buffer. By fixing the depth of the FIFO buffer and manipulating the frequency of the sample generation, which then are passed through the Deep FIFO, we are able to emulate various delays, reaching RTT values of up to 600 ms. 
\begin{table}[hb]
\caption{Experimental configuration}
\label{table}
\centering
\begin{tabular}{|c|c|}
\hline
\textbf{Knobs}& 
\textbf{Meters}\\
\hline
Band& 7 \\
Duplex& FDD\\
Downlink frequency& 2.68 GHz \\
Uplink frequency& 2.56 GHz \\
Bandwidth& 5 MHz\\
TM mode& 1\\
Satellite Attitude& 600 - 35793 km\\
Satellite payload& regenerative/transparent\\
RTT Range& 4 - 480 ms\\
\hline
\multicolumn{2}{p{200pt}}{}\\
\multicolumn{2}{p{200pt}}{*The table indicate the baseline for the Base Station, UE and NTN channel configuration which have been used in this experiment}
\end{tabular}
\label{tab5}
\end{table}

We specifically used in that setup USRPn310 from National Instruments as a software-defined base station and software-defined UE. Further, the NTN channel emulator receive and transmit the signals from/to SDRs using SMA (SubMiniature version A) RF connectors. Accordingly, we assume that the software-defined base station can either be placed at the NTN platform, e.g. in case of a regenerative payload type, or be placed on-ground in case of a transparent payload type. Clearly, the payload type will impact also the RTT in the communication link. In this experiment we assume an NTN altitude ranging between 600 to 35793 km, which results in RTT values between 4 to 480 ms. Table \ref{tab5} shows some of the most relevant configuration parameters. The system uses the LTE band 7, which is a frequency division duplex band (FDD) operating on a downlink frequency 2.68 GHz with an uplink offset of 120 MHz. In LTE the available bandwidths (BW) are 1.4, 5, 10, and 20 MHz. For this experiment we use the 5 MHz BW configuration.

\begin{figure*}[t]
	\centering
	\includegraphics[width=0.9\linewidth,keepaspectratio]{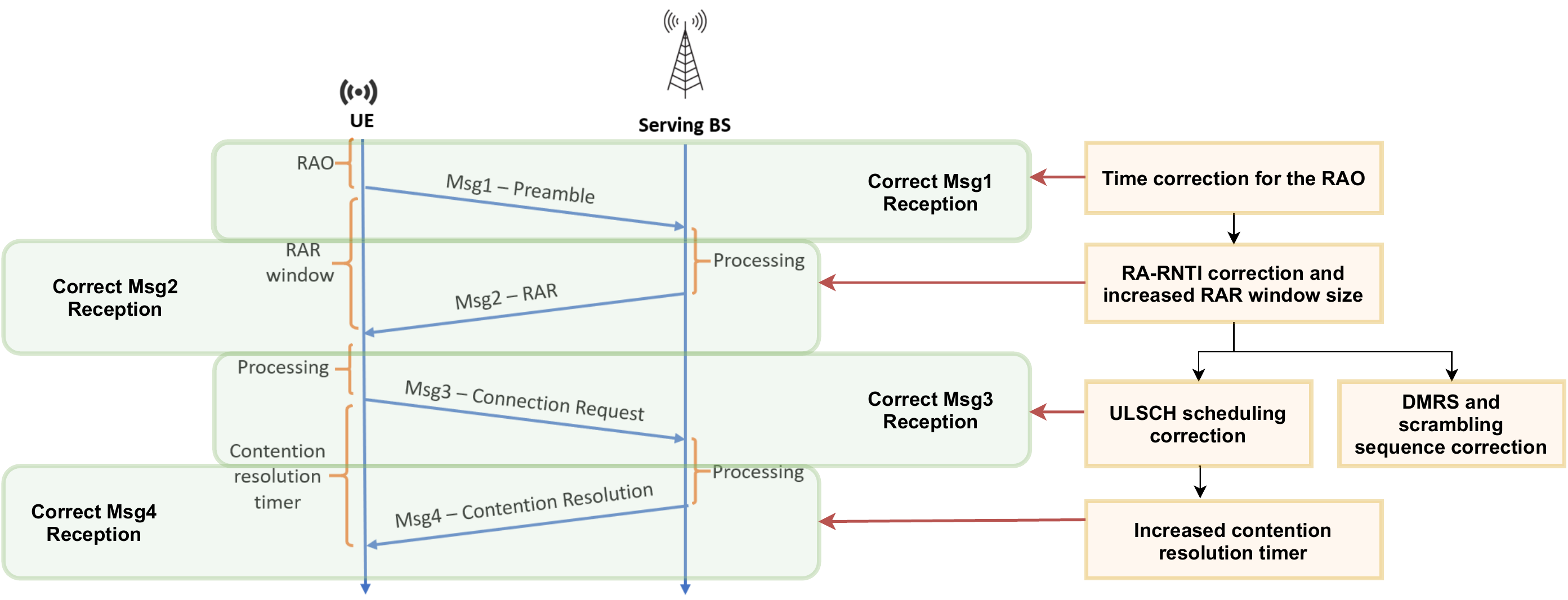}
	\caption{{NTN adaptations in the experiment that led to a step-by-step success of the RA procedure.}}
	\label{fig13}
\end{figure*}
  
\begin{figure}[t]
	\centering
	\includegraphics[width=\columnwidth,keepaspectratio]{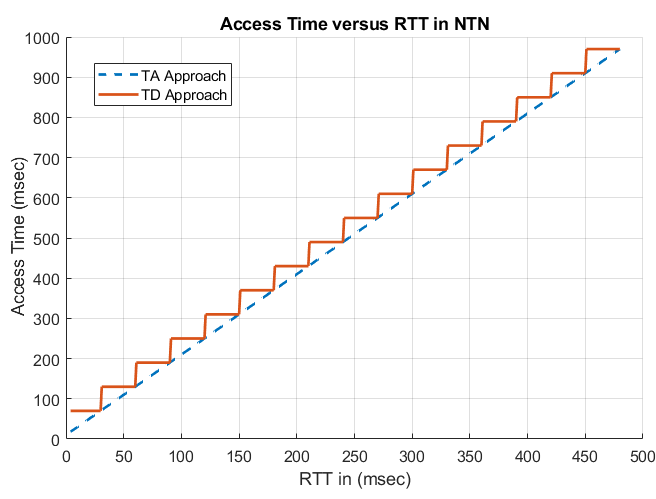}
	\caption{Single User Access Time (no collision)}
	\label{fig14}
\end{figure}

\subsection{Experimental Results}

Through the experimental setup we validate that our proposed solutions are able to counteract the induced challenges by the extreme delays in the communication link, thus leading to a successful RA procedure. Figure \ref{fig13} illustrates the specific adaptations implemented in order to have a successful signal reception, either at UE or BS, for every single step involved in the RA procedure. Such trial-and-error approach enabled us to identify the new challenges not previously treated in the literature, such as the Message 2 withdrawal (Section \ref{sec3b}) and the Message 3 demodulation and decoding problems (Section \ref{sec3e}). Furthermore, we tested that the TA concept, as previously proposed in the literature (although lacking an implementation-wise description), has practical limitations if applied only at a sample-level, leading us to propose a SF-level TA. While this was able to solve the RAO mismatch between the UE and the BS (Section \ref{sec3a}), it failed to deal with the Message 3 and HARQ scheduling mismatch (Section \ref{sec3d}). This is the reason why for the TD approach all the NTN adaptations can be applied at the BS side (see Figure \ref{fig11}), whereas for the TA approach we still need the Message 3 and HARQ scheduling mismatch to be resolved at the BS, and not at the UE (see Figure \ref{fig12}). 

In addition to achieving a successful RA procedure, we measure the time required by a single user to access the network under different values of RTTs, illustrated in Figure \ref{fig14}. As it can be seen, the TD approach results in a step-function behaviour because there are discrete values of the RAR window size and contention resolution timers reported by the BS to the UE (see Equation (4)). In our experiment, we add 16 more values (extra 4 bits) to the already existing ones, covering the NTN delays. In the TA approach we assume a perfect estimation of the TA by the UE, thus updating the RAR window size and contention resolution timer accordingly. Please note that we do not consider any extra time for the GNSS estimate since this process can be done in parallel with obtaining PSS/SSS synchronization. If it is done after, as shown in Figure \ref{fig7}, clearly the access time would be impacted. This is something we cannot measure in our experiment.

In a terrestrial-network case, the main delay bottleneck is caused by the processing time at the BS and UE. For LTE, such processing delay is around 4 ms, resulting in an access delay of around 12 ms in ideal channel conditions and without preamble collisions.

In an NTN case, the delay in the communication link becomes a driving factor in deciding the overall time the user needs for acquiring (re-acquiring) access to the serving BS. As it can be observed from Figure \ref{fig14}, the access time for a single user can be as high as 970 msec in the worst case scenario of a GEO satellite with transparent payload. Clearly, these results represent only a lower bound since a single user is utilized for the experiment and represents a case where no collisions occurs. In a scenario where many user will try to access the network simultaneously, the presence of collisions in the RA procedure will naturally lead to a further increase in the user access times. This is something we are not able to measure in the current experimental setup due to hardware limitations. Nevertheless, the obtained results are a big step forward towards NTN systems. Through this work, we are able to demonstrate a successful RA procedure even in extreme values of delays experienced over a GEO satellite and measure the single user access time, which will act as a lower bound for future work in this direction. 

%%%%%%%%%%%%%%%%%%%%%%%%%%%%%%%%%%%%%%%%%%%%%%%%%%%%%%%%%%%%%%%%%%%%%%%%%%%%%%%%%%%%%%

\section{Conclusions}\label{sec6}

In this paper, we considered the RA procedure, which is a highly crucial mechanism in several standards, such as LTE, NB-IoT and NR, mainly utilized by the users to obtain uplink synchronization in the network. We analyzed the challenges imposed by the increased delay in an NTN channel into the RA procedure, and came up with novel practical solutions for the challenges identified. To test the proposed solutions, we designed a testbed based on  OAI  implementation  for  the  3GPP  users  and  base station, and HW implementation for the NTN channel emulating the signal propagation delay. The laboratory test-bed built in this work enabled us to identify novel challenges not previously treated in the literature, demonstrate a successful RA procedure over NTN, and measure the single-user access times for various values of RTTs. The work shown in this paper increases the technology readiness level (TRL) of NTN-based cellular systems, demonstrating over a laboratory environment a successful RA procedure even in extreme values of delays experienced in a GEO satellite. 

In the future work, we aim to improve the capabilities of our experimental setup in order to be able to test a scenario where multiple user will have to access the network at the same time. Doing so will increase the probability of collisions, enabling us to obtain further results regarding the user access time as a function of the number of users in an NTN network. Furthermore, we will analyse additional challenges in the RA procedure caused by another crucial impairment coming from the NTN platform movement, which is the Doppler shift. Last but not least, solutions where major changes are required in the standard will be implemented and tested as well, such as developing new preamble formats with larger CP lengths able to counterbalance the NTN coverage limitations highlighted in Section \ref{sec4d}.

%\section*{Acknowledgements}

%This work is supported by the Fond National de la Recherche Luxembourg (FNR), under the SATIOT project (Communication algorithms for an end-to-end satellite-IoT).

\bibliographystyle{IEEEtran}
\bibliography{IEEEabrv,access}

\begin{IEEEbiography}[{\includegraphics[width=1in,height=1.25in,clip,keepaspectratio]{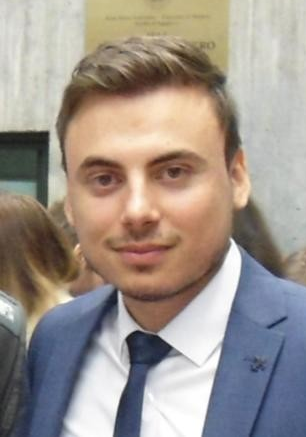}}]{Oltjon Kodheli} received the master’s degree (cum laude) in Electronic Engineering from the University of Bologna, Italy, in 2016. From 2017 to 2018, he was with the Department of Electrical and Information Engineering (DEI), University of Bologna, for performing research related to 5G Radio Access Network. Since 2018, he has been a Doctoral Researcher with the Interdisciplinary Centre for Security, Reliability and Trust, University of Luxembourg. His research activity is mainly focused on designing techniques for 5G IoT satellite-terrestrial integrated systems. He holds a grant for his Ph.D. project received from Luxembourg National Research Fund (FNR), under Industrial Fellowship Scheme, with industrial partner SES S.A.
\end{IEEEbiography}

\begin{IEEEbiography}{Abdelrahman Astro} received a B.Sc degree in Electrical Engineering from the Canadian International College, Egypt, in 2017.  He is currently an RnD Specialist at Interdisciplinary Centre for Security, Reliability and Trust, University of Luxembourg. His research interests include software-defined radios (SDR), wireless communication systems, real-time systems, and systems architecture.
\end{IEEEbiography}

\begin{IEEEbiography}[{\includegraphics[width=1in,height=1.25in,clip,keepaspectratio]{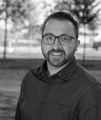}}]{Jorge Querol} (S’13–M’18) was born in Forcall, Castelló, Spain, in 1987. He received the B.Sc. (+5) degree in telecommunication engineering, the M.Sc. degree in electronics engineering, the M.Sc. degree in photonics, and the Ph.D. degree (Cum Laude) in signal processing and communications from the Universitat Politècnica de Catalunya - BarcelonaTech (UPC), Barcelona, Spain, in 2011, 2012, 2013 and 2018 respectively.
His Ph.D. thesis was devoted to the development of novel anti-jamming and counter-interference systems for Global Navigation Satellite Systems (GNSS), GNSS-Reflectometry, and microwave radiometry. One of his outstanding achievements was the development of a real-time standalone pre-correlation mitigation system for GNSS, named FENIX, in a customized Software Defined Radio (SDR) platform. FENIX was patented, licensed and commercialized by MITIC Solutions, a UPC spin-off company.
Since 2018, he is with the SIGCOM research group of the Interdisciplinary Centre for Security, Reliability, and Trust (SnT) of the University of Luxembourg, Luxembourg and head of the Satellite Communications Laboratory. He is involved in several ESA and Luxembourgish national research projects dealing with signal processing and satellite communications. His research interests include SDR, real-time signal processing, satellite communications, 5G non-terrestrial networks, satellite navigation, and remote sensing.
He received the best academic record award of the year in Electronics Engineering at UPC in 2012, the first prize of the European Satellite Navigation Competition (ESNC) Barcelona Challenge from the European GNSS Agency (GSA) in 2015, the best innovative project of the Market Assessment Program (MAP) of EADA business school in 2016, the award Isabel P. Trabal from Fundació Caixa d’Enginyers for its quality research during his Ph.D. in 2017, and the best Ph.D. thesis award in remote sensing in Spain from the IEEE Geoscience and Remote Sensing (GRSS) Spanish Chapter in 2019.

\end{IEEEbiography}

\begin{IEEEbiography}[{\includegraphics[width=1in,height=1.25in,clip,keepaspectratio]{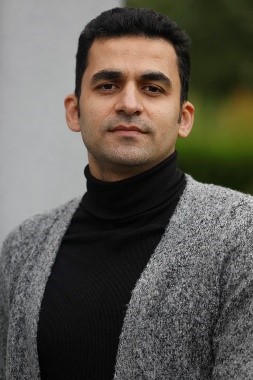}}]{Mohammad Gholamian} received his B.Sc. degree in electrical engineering at the Electrical Engineering department of Hakim Sabzevari University, Mashhad (Iran), in 2007. He has worked for many years in Research Centers, and more specifically for the implementation of signal processing and communication algorithms on different FPGA platforms and costume boards. His research interests are related to digital signal processing for wireless communication systems, DSP/FPGA implementation of signal processing algorithms, Software Defined Radios (SDR) and embedded systems. since 2020 he has joined to the Signal Processing and Communications research group, SIGCOM, at the Interdisciplinary Centre for Security, Reliability and Trust, University of Luxembourg.
\end{IEEEbiography}

\begin{IEEEbiography}[{\includegraphics[width=1in,height=1.25in,clip,keepaspectratio]{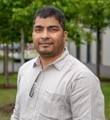}}]{Sumit Kumar}
(S'14-M'19) received his
Bachelor of Technology and Master of
Science in Electronics \& Communication
Engineering from Gurukula Kangri
University, Haridwar, India (2008) and
the International Institute of Information
Technology, Hyderabad, India (2014),
respectively, and his PhD degree from
Eurecom (France) in 2019. His research interests are in wireless
communication, interference management and software defined
radio.
\end{IEEEbiography}

\begin{IEEEbiography}[{\includegraphics[width=1in,height=1.25in,clip,keepaspectratio]{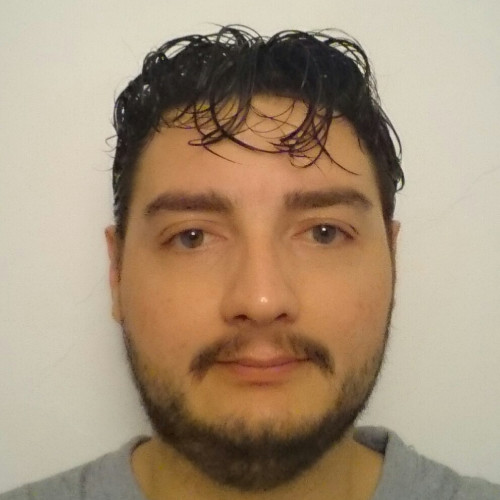}}]{Nicola Maturo} received his M.S. Degree in Electronic Engineering (cum laude) from the Polytechnic University of Marche, Ancona (Italy), in 2012 and his Ph.D. on Telecommunication Engineering in the same University, in 2015. 
From January 2016 to July 2017 he was a Post-Doctoral Researcher at the Department of Information Engineering of the Polytechnic University of Marche, where he worked on error correcting coding techniques under some ESA research projects. From November 2015 to May 2016 he was consultant for Deimos Engenharia (Lisbon) working on spectral estimation algorithms and anti-jamming techniques. Since August 2017 to July 2021 he was Research Scientist at the Interdisciplinary Centre for Security, Reliability and Trust of the University of Luxembourg mainly working in the satellite communication domain. He is a member of IEEE since 2013 and of the CCSDS Coding and Synchronization Working Group since 2015.
\end{IEEEbiography}

\begin{IEEEbiography}[{\includegraphics[width=1in,height=1.25in,clip,keepaspectratio]{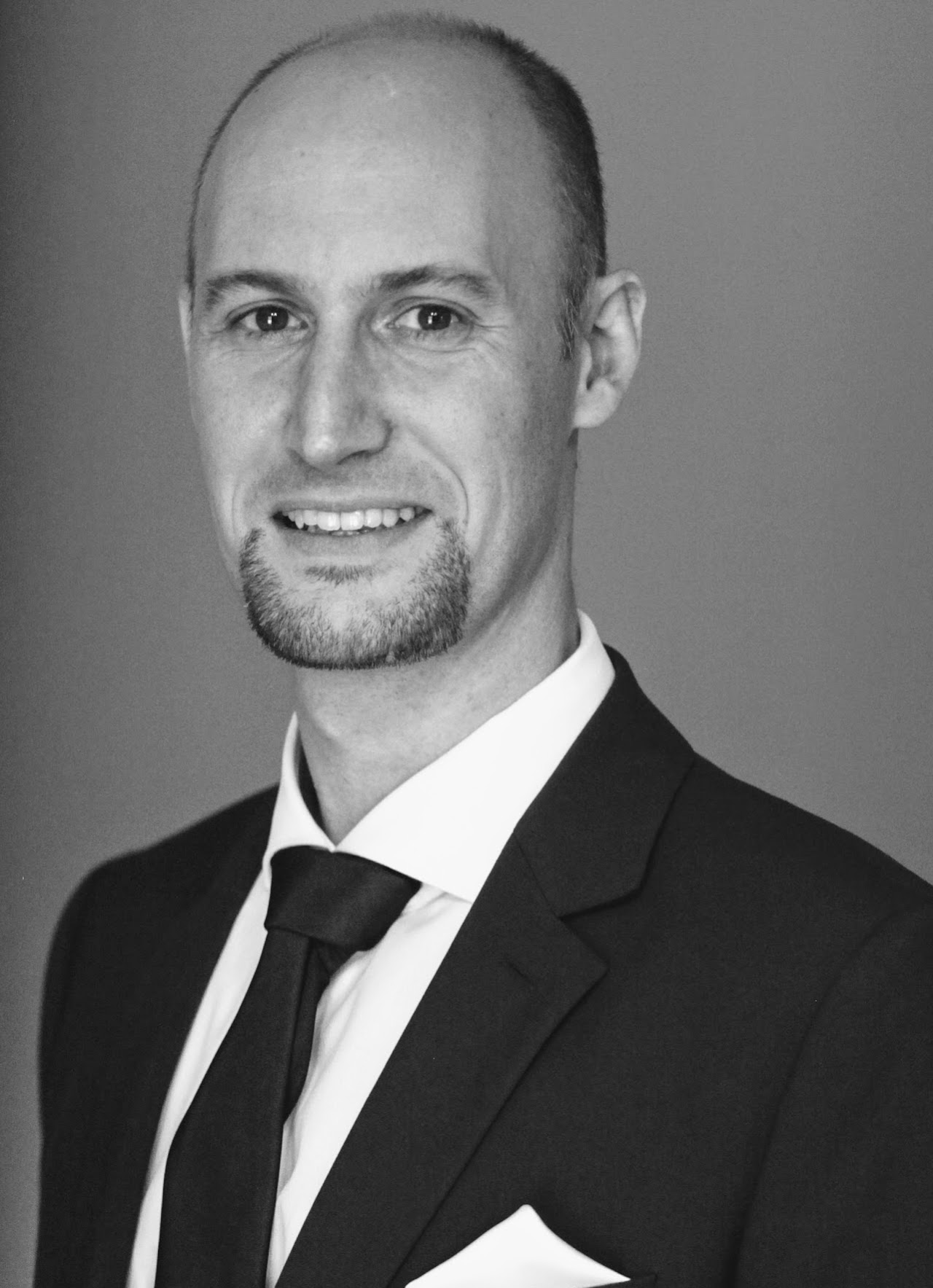}}]{Symeon Chatzinotas} is currently Full Professor / Chief Scientist I and Head of the SIGCOM Research Group at SnT, University of Luxembourg. He is coordinating the research activities on communications and networking, acting as a PI for more than 20 projects and main representative for 3GPP, ETSI, DVB.
In the past, he has been a Visiting Professor at the University of Parma, Italy, lecturing on “5G Wireless Networks”. He was involved in numerous R\&D projects for NCSR Demokritos, CERTH Hellas and CCSR, University of Surrey.
He was the co-recipient of the 2014 IEEE Distinguished Contributions to Satellite Communications Award and Best Paper Awards at EURASIP JWCN, CROWNCOM, ICSSC. He has (co-)authored more than 450 technical papers in refereed international journals, conferences and scientific books.
He is currently in the editorial board of the IEEE Transactions on Communications, IEEE Open Journal of Vehicular Technology and the International Journal of Satellite Communications and Networking.
\end{IEEEbiography}

\EOD

\end{document}